\documentclass[11pt]{article}
\usepackage{mydef2col}
\usepackage{markArticle}
\usepackage{graphicx}

\providecommand{\Kmat}{\mathbf{K}}
\providecommand{\EE}{\mathbb{E}}
\ifdefined\theorem\else\newtheorem{theorem}[proposition]{Theorem}\fi

\title{Spectral Pricing of Corridor Products under Layered Local Volatility}

\shorttitle{Spectral Pricing of Corridor Products}

\author{
\authorstyle{Andrey Itkin\textsuperscript{1}\thanks{e-mail: \url{aitkin@nyu.edu}} \, and
Dmitry Muravey\textsuperscript{2}\thanks{e-mail: \url{d.muravey87@gmail.com}}
}
\newline\newline
\textsuperscript{1}
\institution{FRE Department, Tandon School of Engineering, New York University, USA} \\
\textsuperscript{2}
\institution{Abu Dhabi Investment Authority, UAE}\\
}

\graphicspath{{figs/}}

\date{\vspace{-2.\baselineskip}\today}

\begin{document}

\maketitle

\lettrineabstract{We price corridor products under a $N$-layered local volatility, piecewise constant in the price and in time, with knock-out barriers. A spatial scale transformation and the GIT method reduce the killed Green's function to a symmetric tridiagonal system of fixed size $N-1$. With equal scaled layer widths the spectrum follows from one tridiagonal eigen-decomposition together with the explicit eigenfunctions that vanish at every interface. For general scaled layer widths an exact counting theorem locates every eigenvalue. Consecutive time intervals of constant volatility are connected by a matrix that re-expands the solution in the eigenbasis of the preceding interval, and the result converges exponentially in the number of retained terms in the eigenfunction expansion. Allowing the killing rate to vary between layers preserves this construction. On a single time interval, the corridor accrual distribution is then recovered by inverting the time and Fourier transforms, without introducing an additional state variable. Exact representations for Greeks and layer vegas follow by differentiation. In our benchmarks the method agrees with independent calculations near rounding error and keeps full accuracy at volatility interfaces, where the tested uniform finite-difference (FD) grids lose one order of convergence. The method computes the accrual law about sixty times faster than the tested FD scheme that includes accrued time as a second state variable.}

%\medskip
%\noindent\textbf{Keywords:} corridor options; range accrual; corridor variance swap; occupation time; piecewise-constant local volatility; heat potentials; Volterra equations; tridiagonal spectra; theta functions.
%
%\medskip
%\noindent\textbf{JEL:} C02, C63, G12, G13.
%
%\medskip
%\noindent\textbf{MSC:} 35K05, 35Q91, 42A38, 44A10, 45D05, 60J60, 91G20.

%%%%%%%%%%%%%%%%%%%%%%%%%%%%%%%%%%%%%%%%%%%%%%%%%%%%%%%%%%%%%%%%%%%
\section{Introduction}\label{sec:intro}

Range accruals, corridor swaps and corridor variance swaps all pay a functional of the time the underlying spends inside a band. A range accrual note pays a coupon proportional to the fraction of fixings at which the reference rate lies inside the corridor. A corridor swap exchanges such accruals over a schedule of reset dates. A corridor variance swap accumulates realized variance only while the underlying is inside the corridor, \cite{CarrLewis2004}. Knock-out versions of corridor variance swaps are analyzed in \cite{ahallal2018knocking}.

When the payoff is linear in the accrued time, only the expected occupation time matters. Many traded variants are not linear. Capped and floored accruals, digital range accruals that pay only when the accrued fraction exceeds a threshold, and options on the accrual itself all depend on the law of the occupation time, not only on its mean. A direct finite-difference (FD) scheme handles such payoffs by adding the running accrual as a second state variable, which turns a one-dimensional pricing problem into a two-dimensional one. The method developed here obtains the law of the accrual without that extra dimension. A one-dimensional FD solver can also be applied after transformation in the accrual variable. Our comparison uses the direct augmented-state FD scheme.

In this paper we choose to work with a local volatility that is piecewise constant in the price, for two reasons. First, the model is parsimonious. It has one volatility per layer and calibrates directly to a small number of quotes. Second, it is the natural step approximation of a Dupire local volatility surface, \cite{Dupire:94}. The representations below therefore give the exact prices of the discretized model. Their numerical evaluation requires truncation or inversion, and the error of the step approximation is a separate question about the model. Corridor edges and barriers fit the structure without further assumptions, because each of them can be made a breakpoint of the volatility.

For practical use the volatility must also move with time, at least piecewise. We therefore also treat a tiled local volatility, piecewise constant in the price and in time, the form calibrated in \cite{liptonSepp2011}. Within each time interval on which the volatility is constant, which we call a \emph{time bucket}, the analysis below is exact. Consecutive time buckets are connected by a matrix that re-expands the solution in the eigenbasis of the preceding bucket. For fixed positive bucket lengths, this step converges exponentially in the number of modes \footnote{Modes are eigenfunctions of the spatial operator satisfying the barrier and interface conditions. These include eigenfunctions that are nonzero at some interfaces and those that vanish at every interface. The number of retained modes is the number of terms kept in the eigenfunction expansion.} and in our tests reaches machine precision with about twenty modes per bucket. This extension covers terminal payoffs and expected running rewards, which are paid continuously in time, such as accruals and corridor variance. The full law of the accrual is developed for a single time bucket.

The method is intended for routine use in practice, not only for analysis. A single-bucket price takes about two milliseconds on a standard laptop, with independent pricing methods agreeing near rounding error in the reported tests. We call a Crank--Nicolson scheme fitted when its grid has a node on every volatility interface and the squared volatility at such a node is taken as the harmonic mean of its values in the two adjacent grid cells, weighted by the cell widths. The fitted scheme needs 120 milliseconds to reach an error of $4 \times 10^{-6}$. A scheme on a uniform grid that samples the volatility at the nodes drops to first order in the presence of volatility interfaces, which is where corridor payoffs are most sensitive.

The Greeks come from the same eigen-data. Delta, gamma and theta are closed form, gamma is returned on both sides of each interface where it can jump, and all ten layer vegas of a ten-layer model take twenty milliseconds using complex-step differentiation. The resulting vega Jacobian calibrates the layer volatilities to vanilla quotes in five Newton steps and under one second.

For range accruals with a payoff nonlinear in the accrued time, such as digital and capped accruals, the single-bucket law of the accrual comes out about sixty times faster than from the tested FD scheme that carries the accrual as a second state variable, with an error more than twenty times smaller in the reported comparison of the cumulative distribution function (CDF). With volatility piecewise constant in time, a surface of ten layers and twelve monthly buckets prices a terminal payoff in about a tenth of a second, and all 120 tile vegas take four seconds.

The analytic literature on occupation-time derivatives rests almost entirely on homogeneous dynamics. Early work priced quantile options using the occupation-time distribution of Brownian motion with drift, \cite{Akahori1995, Dassios1995}. General occupation-time claims under geometric Brownian motion were subsequently addressed via Feynman--Kac methods, \cite{Hugonnier1999}. Closed-form solutions are available for step options, where the knock-out rate is proportional to the time spent below a barrier, \cite{Linetsky1999}. Furthermore, the law of the time spent in a band and the pricing of corresponding corridor options were derived in \cite{Fusai2000}. Discrete monitoring was treated in \cite{FusaiTagliani2001}, while corridor variance swaps were introduced in \cite{CarrLewis2004}. More recently, occupation-time pricing was extended to the Kou jump-diffusion model, \cite{CaiChenWan2010}.

In several of these models, the coefficients are constant after a suitable change of variable, such as log price under geometric Brownian motion. Consequently, the Feynman--Kac equation features constant coefficients on each side of the corridor edge, allowing solutions via elementary or special functions. However, general state-dependent volatility removes this constant-coefficient structure. Our model restores analytic tractability by making the state dependence piecewise constant. The corridor edges then function as interfaces, identical in nature to the volatility breakpoints.

Diffusions with piecewise-constant volatility have an established history. The two-valued case on the real line is known as oscillating Brownian motion, \cite{KeilsonWellner1978}. Pricing under a two-valued local volatility has been achieved using the density of skew Brownian motion and its functionals, \cite{GairatShcherbakov2017}. Skew Brownian motion has also served directly as a pricing model, \cite{CornsSatchell2007}. Notably, two-valued local volatility can reproduce extreme at-the-money skews, \cite{Pigato2019}. The joint density approach was recently utilized to analyze this phenomenon under discontinuous local volatility, \cite{GairatShcherbakov2024}. This same methodology was previously employed by the authors to model skew Brownian motion with dry friction, \cite{GairatShcherbakov2022}.

On the statistical side, parameters of oscillating Brownian motion were estimated in \cite{Lejay2018}, and a threshold local volatility model was fitted to historical data in \cite{LejayPigato2019}. The convergence rate of the maximum likelihood estimator for the skewness parameter was established in \cite{Lejay2024}. Alternatively, volatility interfaces can be handled via delta-function perturbations, \cite{DecampsDeSchepperGoovaerts2004}. A tiled local volatility, piecewise constant in both space and time, was successfully calibrated using semi-analytic solutions of the forward equation, \cite{liptonSepp2011}. This framework was later extended to a local volatility that is piecewise linear in space. This extension ensures smooth fits of the implied volatility surface while preserving semi-analytic tractability, \cite{ItkinLipton2017}. This trajectory of research, originating from the Dupire and Derman--Kani models, \cite{Dupire:94, DermanKani1998} and extending to piecewise local volatility calibration, is systematically detailed in \cite{ItkinLocalVol}.

Recent analytical progress has further refined these techniques. The Multilayer (ML) method combines the method of heat potentials with the generalized integral transform (GIT) method to reduce diffusion problems on layered domains to systems of Volterra equations, \cite{ItkinLiptonMuravey2022_Finance}. This construction was extended through oscillating integral transforms to multilayer problems with moving interfaces, \cite{ItkinLiptonMuravey2022_PhysicaA}. Our starting point is the layer representation and the associated Volterra system for the interface values developed in these works. Related results on occupation times of one-dimensional and skew diffusions are given in \cite{SalminenStenlund2021, SalminenStenlund2025}.

Despite these advances, a critical gap remains for corridor products. The ML method and its transform extensions deliver the Green's function of a layered strip. For fixed interfaces, we make its spectral construction explicit: equal scaled layer widths (SLW)\footnote{The scaled layer width of a layer is $\sqrt{2}$ times its width divided by its local volatility. Equal SLW therefore means that the layer widths are proportional to the layer volatilities.} reduce the interface modes to one fixed matrix, whereas general scaled layer widths require a parameter-dependent interface matrix. Pricing a nonlinear occupation-time product under a calibrated step volatility requires the law of the relevant occupation functional. It also requires a mapping that transitions the layered analytical price into a driftless local volatility martingale. This paper supplies these objects. The counting formula for general SLW is a specialization of the Wittrick--Williams principle, \cite{CorridorWittrickWilliams1971}.

Corridor variance swaps were formally introduced by \cite{CarrLewis2004}. Their pricing and hedging under more general conditions, including quasi-static super-replication strategies, were further developed by \cite{BurgardTorne2018}. Our approach complements this literature by providing spectral representations and occupation-time transforms through a tridiagonal interface matrix. The standard corridor variance swap measures log-return quadratic variation, \cite{CorridorLee2010}. We distinguish it from price quadratic variation and provide pricing formulas for both, including the extension of their expected rewards to tiled volatility.

Double-barrier options under homogeneous dynamics are classical instruments. They have been priced using image expansions, \cite{KunitomoIkeda1992}, Laplace transforms of exit times, \cite{GemanYor1996}, contour inversion of transforms, \cite{Pelsser2000}, and systematic transform treatments, \cite{Lipton2001}. For smooth scalar diffusions, barrier claims were priced via Sturm--Liouville eigenfunction expansions, \cite{DavydovLinetsky2003}. The layered model presented here fits the Sturm--Liouville framework with transmission conditions. The eigenfunctions are only piecewise smooth and are glued together by these conditions. Our spectral series makes this construction explicit. Under equal SLW, dividing each row by a scalar factor gives a multiple of the identity minus a fixed matrix that does not depend on the Laplace variable. Each off-diagonal entry of this fixed matrix is the probability that the underlying, started at an interface, reaches a given neighboring interface before the other one. The interface modes are then determined by this fixed $(N - 1) \times (N - 1)$ matrix.

The analytical method of this paper is based on the GIT method. The method has previously yielded semi-closed form prices for barrier options under time-dependent Ornstein--Uhlenbeck, CEV, CIR, and Hull--White dynamics. The approach was later extended to the time-dependent Heston and SABR models, \cite{CarrItkinMuraveyHeston,ItkinMuravey2022sabr} and to American options in time-dependent one-factor models, \cite{ItkinMuravey2024jd}. Similarly, the method of heat potentials was applied to first-hitting-time densities, \cite{LiptonKaushansky2020} and to barrier options under stochastic volatility, \cite{LiptonSepp2022}. Both foundational methods are detailed comprehensively in \cite{ItkinLiptonMuraveyBook}.

We introduce three new elements to the ML construction. First, the Laplace-transformed system is reduced to a symmetric tridiagonal matrix. This matrix realizes the Dirichlet-to-Neumann map of the layered strip on its interfaces. Under equal SLW, its row normalization yields the substochastic embedded interface-chain matrix. The interface modes are then determined by a fixed $(N - 1) \times (N - 1)$ matrix. The flux matrix itself is not a transition matrix. Second, we introduce layer-dependent killing rates, which are coefficients of additional linear decay terms in the evolution equation. Each rate produces an exponential weight determined by the time spent in its layer. By the Feynman--Kac formula, the same tridiagonal system then yields transforms of the layer occupation times on survival and at exit. Third, a spatial scale transformation ensures the layered analytics directly correspond to a driftless local volatility martingale, which is the exact requirement for corridor pricing.

A separate line of literature prices barrier and occupation-time claims using continuous-time Markov chain approximations, \cite{MijatovicPistorius2011, CuiKirkbyNguyen}. In that framework, the diffusion is replaced by a chain on a grid, and its tridiagonal generator is exponentiated. Our work is complementary to this approach. In the Markov chain literature, the tridiagonal matrix is a discretization whose dimension must grow to achieve desired accuracy. In our framework, the tridiagonal matrix represents the exact spectral data of the continuum model. Its dimension is strictly fixed at $N - 1$ by the model itself, and time is never discretized within a bucket. Numerical evaluation still requires spectral truncation or transform quadrature. These hitting probabilities describe only the order in which the underlying visits the interfaces. They do not replace the diffusion by a finite-state process. This positions our model as a natural, exact benchmark for the Markov chain approximation literature.

This paper bridges the gap between piecewise diffusion analytics and the practical requirements of pricing corridor products. Our main contributions are sevenfold:

\begin{enumerate}[itemsep=10pt]
\item \emph{Scale reduction:} We introduce a mapping that identifies the layered divergence-form problem with the double-barrier pricing problem for a genuine local volatility martingale.

\item \emph{Closed-form Green's function:} We derive an exact representation of the killed Green's function\footnote{In probability theory, the killed Green's function (or the Green's function of a killed process) corresponds to the transition density or resolvent kernel of a stochastic process (such as Brownian motion or a diffusion) that is instantaneously terminated ("killed") upon exiting a domain or hitting a boundary, or via an independent exponential clock defined by a killing potential. In mathematical physics, this probabilistic formulation is equivalent to the Green's function of the governing differential operator (e.g., the Laplacian or Fokker-Planck operator) on a finite interval or bounded domain subject to homogeneous Dirichlet boundary conditions, where the boundaries act as absorbing barriers.} via a Volterra system whose Laplace transform is symmetrically tridiagonal.

\item \emph{Matrix identification and spectral simplicity:} Under equal SLW, we show that the fixed matrix consists of the hitting probabilities between neighboring interfaces. Its real, simple eigenvalues determine the interface modes. Modes vanishing at every interface complete the diffusion spectrum. For unequal SLW, an exact counting theorem built on the same matrix locates every eigenvalue.

\item \emph{Joint law of occupation times:} We show that layer-dependent killing rates, which generate the occupation times through the Feynman--Kac formula, preserve the tridiagonal structure. This construction gives transforms of occupation on survival and of exit time and occupation at exit. On a single time bucket, time inversion followed by one Fourier inversion gives the law of the corridor accrual, and a lognormal local volatility enters the same system through a gauge transformation and a particular choice of the killing rates.

\item \emph{Exact Greeks:} Delta, gamma and theta follow from the eigen-series in closed form. Layer vegas follow from a Hellmann--Feynman formula for the eigenvalues, \cite{Feynman1939} and from term-by-term differentiation of the series.

\item \emph{Speed and accuracy, measured:} The independent pricing methods agree near rounding error in the reported tests and keep full accuracy at volatility interfaces, where the tested uniform FD grids lose an order. Short-time and long-time representations provide alternatives across maturities. The benchmarks compare with Crank--Nicolson for prices and with a direct augmented-state implicit-Euler FD scheme for the accrual law.

\item \emph{Volatility piecewise constant in time:} A matrix that re-expands the solution from the eigenbasis of one time bucket into that of the next extends terminal-payoff prices, Greeks, expected accruals and expected corridor variance to tiled local volatility, with exponential convergence for positive bucket lengths and an explicit norm bound on the spectral truncation error. Both price and log-return quadratic variation are covered.

\end{enumerate}

The rest of the paper is organized as follows. \cref{sec:model} sets up the tiled model, the scale transformation and the three primitives to which every product reduces. \cref{sec:potentials,sec:laplace,sec:equal,sec:general} analyze a single time bucket. \cref{sec:potentials} derives the Volterra system for the interface densities from the GIT representation, and \cref{sec:laplace} reduces it to a symmetric tridiagonal system in the Laplace domain. \cref{sec:equal} inverts the system in closed form when the SLW are equal, and \cref{sec:general} treats general SLW through an exact eigenvalue count. \cref{sec:barrier} prices barrier and no-touch claims, \cref{sec:corridor} prices corridor products and derives the law of the accrual, and \cref{sec:greeks} gives the Greeks. \cref{sec:tiled} joins the time buckets. \cref{sec:numerics} reports the implementation and the benchmarks, and \cref{sec:conclusion} concludes. The appendices contain the lognormal and drifted variants, the derivation of the GIT representation of a layer, the algebra of layer-dependent killing and the eigen-solver.

%=======================================================================
\section{The model and the pricing problem}\label{sec:model}

Under the pricing measure, with $W$ a standard Brownian motion with respect to a filtration $(\mathcal F_t)_{t \ge 0}$, the underlying is a driftless martingale,
\begin{equation}\label{eq:S}
d S_t = \vs(t, S_t)\,d W_t, \qquad S_0 = S_*,
\end{equation}
on a strip between two barriers $B_- < B_+$. The volatility is piecewise constant in the price and in time. Let $B_- = Y_0 < Y_1 < \dots < Y_N = B_+$ partition the strip into layers $(Y_{i - 1}, Y_i)$ of width $L_i := Y_i - Y_{i - 1}$, and let $0 = t_0 < t_1 < \dots < t_{\bar K} = T$ partition the maturity. We call $[t_{k - 1}, t_k)$ the $k$-th time bucket and the rectangle $(Y_{i - 1}, Y_i) \times [t_{k - 1}, t_k)$ a tile. On the $k$-th bucket
\begin{equation}\label{eq:vspwc}
\vs(t, y) = \sum_{i = 1}^{N} \vs_{i, k}\, \ind_{\{Y_{i - 1} \le y < Y_i\}}, \qquad t \in [t_{k - 1}, t_k),
\end{equation}
with $\vs_{i, k} > 0$. This is the tiled local volatility of \cite{liptonSepp2011}. The breakpoints, and hence the barriers, are the same in every bucket. The half-open convention assigns a positive volatility to each interior breakpoint. Any positive choice there defines the same non-sticky diffusion. The process is stopped on reaching a barrier.

The analysis first treats a single bucket, on which the volatility depends on the price only. Within the bucket we drop the bucket index and write $\vs_i$ for the volatility of layer $i$ and $\vs(y)$ for the volatility at the price $y$. Write
\begin{equation}
\tau := \inf\{ t \ge 0 : S_t \notin (B_-, B_+) \}
\end{equation}
for the exit time from the strip.

The barriers may be genuine contractual barriers or, for an unbounded underlying, artificial cut-offs placed far enough out that the truncation error is below the target tolerance. The error is easy to bound. Let $D := \min(S_* - B_-, \, B_+ - S_*)$ and $\vs_{\max} := \max_{i, k} \vs_{i, k}$. Up to the exit time, $S_t - S_*$ is a Brownian motion run at the clock $\langle S \rangle_t \le \vs_{\max}^2 t$ by the Dambis--Dubins--Schwarz theorem, so the reflection principle gives
\begin{equation}\label{eq:trunc}
\Prb(\tau \le T) \le \Prb \Big( \sup_{u \le \vs_{\max}^2 T} |W_u| \ge D \Big)
  \le 2\,\mathrm{erfc}\! \left( \frac{D}{\vs_{\max} \sqrt{2T}} \right).
\end{equation}
For a constant interest rate $r \ge 0$, a claim whose payoff is bounded by $\bar P$ changes by at most $\bar P$ times this bound when the cut-offs are removed. Unbounded payoffs require an additional growth estimate. Suppose the model is extended beyond the cut-offs with $\vs\le\vs_{\max}$, and let $P$ be a Call or Put payoff, hence nonnegative and $1$-Lipschitz. Conditional on an exit at time $\tau\le T$, the martingale increment satisfies $\EE[|S_T-S_\tau|\mid\mathcal F_\tau]\le\vs_{\max}\sqrt{T-\tau}$. Therefore the price $V_{\rm full}$ in the extended model and the price $V_{\rm killed}$ of the claim knocked out at the cut-offs satisfy
\begin{equation}\label{eq:truncvanilla}
0\le V_{\rm full}-V_{\rm killed}\le e^{-rT}\bigl(\max\{P(B_-),P(B_+)\}+\vs_{\max}\sqrt T\bigr)\Prb(\tau\le T).
\end{equation}
The bound is Gaussian in the distance to the cut-off, which is the relevant statement at a fixed maturity. The leading Dirichlet mode governs the decay in $T$ for fixed barriers, which is a different question.

The driftless specification is a modeling assumption. It applies to a spot price with zero cost of carry, $r = d$ with $d$ the dividend yield, or to a forward price of fixed maturity under deterministic interest rates and dividend yields. In the forward interpretation, the volatility breakpoints, knock-out barriers and corridor edges are specified and fixed in the forward coordinate. A level $Y$ fixed in the spot coordinate becomes $Y e^{(r-d)(T-t)}$ in the forward coordinate. Thus, nonzero cost of carry generally produces moving interfaces and contractual boundaries, which are outside the fixed-interface construction developed here. Discounting is treated separately and does not require $r = 0$.

A lognormal local volatility, and the role of a drift, are treated in \cref{app:drift}. Extending the method to moving interfaces and contractual boundaries, including those arising when spot-defined contracts are expressed in forward coordinates with nonzero cost of carry, remains for future work.

\subsection{Reduction to divergence form by a scale transformation}

This is the step that connects the analytic problem below to a martingale model.
\begin{definition}[scale coordinates]
Set
\begin{equation}\label{eq:scale}
\Phi(y) := \sqrt{2} \int_{Y_0}^{y} \frac{d u}{\vs^2(u)}, \qquad x := \Phi(y), \qquad X_t := \Phi(S_t),
\end{equation}
and let
\begin{equation}\label{eq:scaledpars}
y_i := \Phi(Y_i), \qquad \ell_i := y_i - y_{i - 1} =  \frac{\sqrt{2}\,L_i}{\vs_i^{2}}, \qquad \sigma_i := \frac{1}{\vs_i}, \qquad
\omega_i := \frac{\ell_i}{\sigma_i} = \frac{\sqrt{2}\,L_i}{\vs_i}.
\end{equation}
We call $\omega_i$ the scaled layer width (SLW) of layer $i$.
\end{definition}

\begin{proposition}[scale reduction]\label{prop:scale}
Let $u(\theta, y)$ solve the backward pricing problem in the time to maturity
$\theta$,
\[
  u_\theta = \tfrac 12 \vs^2(y)\, u_{yy},
  \qquad y \in (B_-, B_+),
\]
with $u$ and $u_y$ continuous across every $Y_i$ and homogeneous Dirichlet
data at $B_\pm$. Then $v(\theta, x) := u(\theta, \Phi^{-1}(x))$ solves
\begin{equation}\label{eq:pde}
\sigma^2(x)\, v_{xx} = v_\theta, \qquad x \in (y_0, y_N), \qquad
\sigma(x) = \sum_{i = 1}^N \sigma_i \ind_{\{y_{i - 1} < x < y_i\}},
\end{equation}
with $v$ and $\sigma^2(x) v_x$ continuous across every $y_i$, and conversely.
Moreover $\Phi$ is a strictly increasing bijection and involves no time
change, so
\begin{equation}\label{eq:occinvariance}
\tau = \inf\{t : X_t \notin (y_0, y_N)\}, \qquad \int_0^t \ind_{\{S_s \in C\}}\,d s
= \int_0^t \ind_{\{X_s \in \Phi(C)\}}\,ds \quad \text{for every Borel } C.
\end{equation}
\end{proposition}

\begin{proof}
Inside a layer $\Phi$ is affine, so $u_{yy} = v_{xx} \Phi'^2$ and $\tfrac 12 \vs^2 \Phi'^2 = \tfrac 12 \vs^2 \cdot 2/\vs^4 = 1/\vs^2 = \sigma^2$. For the interface condition, $u_y = v_x \Phi' = \sqrt 2\, v_x/\vs^2 = \sqrt 2\, \sigma^2 v_x$, so $u_y$ is continuous if and only if $\sigma^2 v_x$ is. The last claim holds because $\Phi$ acts on space only.
\end{proof}

\begin{myremark}[Auxiliary diffusion and price process] \label{rem:whatisX}
The auxiliary process $X$ is a multiskew diffusion with divergence-form generator $\mathcal{A} = \partial_x\! \left( \sigma^2(x) \partial_x \cdot \right)$, scale density $s'(x) = 1/\sigma^2(x)$, and speed measure $\mathfrak m(dx) = dx$. Unless the layer volatilities are all equal, its scale function is not affine and $X$ is not a martingale. The martingale price process is $S$. Pricing statements are expressed in the price coordinate $y$, whereas the analysis is conducted in the auxiliary coordinate $x$. The scale transformation \eqref{eq:scale} relates these coordinates and makes the distinction between the two processes explicit.
\end{myremark}

\begin{myremark}[Normalization]
Under the convention adopted in \eqref{eq:pde}, a Brownian motion with generator $\sigma^2 \partial_{xx}$ has variance $2\sigma^2 t$. The standard convention $\tfrac12 \bar\sigma^2 \partial_{xx}$ is recovered by setting $\sigma = \bar\sigma/\sqrt{2}$. We retain the normalization in \eqref{eq:pde} throughout.
\end{myremark}

\begin{myremark}[Equal SLW in price coordinates] \label{rem:transit}
The equal SLW condition $\omega_i \equiv \omega$ is equivalent to $L_i/\vs_i = \text{const}$ in the price coordinate. Thus, the layer widths are proportional to the local volatility. When the layer boundaries can be chosen freely, as in the discretization of a Dupire surface, this condition provides a natural choice that can be imposed at negligible additional cost.
\end{myremark}

\subsection{Three primitives}

The pricing representations below are built from three basic objects.
\begin{enumerate}[itemsep=10pt]
\item[(P1)] The killed density $U(t, x; x_0)$, $x_0 = \Phi(S_*)$, i.e.\ the Green's function of \eqref{eq:pde} with Dirichlet ends.

\item[(P2)] The boundary fluxes $\sigma^2 U_x$ at $y_0$ and $y_N$, i.e.\ the
first-passage densities to each barrier.

\item[(P3)] The layer occupation vector
\begin{equation}\label{eq:occvec}
\Lam(t) = \big( \Lambda_1(t), \dots, \Lambda_N(t) \big), \qquad
\Lambda_i(t) := \int_0^{t \wedge \tau} \ind_{\{ X_s \in (y_{i - 1}, y_i)\}}\,d s.
\end{equation}

\end{enumerate}
A claim with payoff $P$ at $T$, knocked out at either barrier, is worth
\begin{equation}\label{eq:pricing}
V(S_*) = e^{-rT} \int_{y_0}^{y_N} P \big( \Phi^{-1}(x) \big)\, U(T, x; x_0)\,d x.
\end{equation}

If the claim also pays a rebate $R_-$ or $R_+$ at the first passage to $B_-$ or
$B_+$ before $T$, add
\begin{equation}\label{eq:rebate}
R_- \int_0^T e^{-rt}\, p_-(t)\, d t + R_+ \int_0^T e^{-rt}\, p_+(t)\, d t,
\qquad p_-(t) := \sigma_1^2\,\partial_x U(t, y_0 +), \quad p_+(t) := -\sigma_N^2\,\partial_x U(t, y_N -).
\end{equation}
These are the first-passage densities of (P2). Both are nonnegative, because $U$ vanishes at the barriers and is positive inside. They are the outflow rates through the two ends. By \cref{rem:whatisX}, $U$ solves $\partial_t U = \partial_x(\sigma^2 \partial_x U)$, and integrating over $(y_0, y_N)$ gives $\frac{d}{d t} \int U\,d x = -p_-(t) - p_+(t)$.

%=======================================================================
\section{The killed Green's function via the GIT method} \label{sec:potentials}

The function $U$ solves the following partial differential equation (PDE) in divergence form
\begin{align}
\partial_x \left( \sigma^2(x) \partial_x U(t, x) \right) &= \partial_t U(t, x), \qquad (x, t) \in [y_0, y_N] \times \mathbb{R}_+, \\
U(t, y_0) &= U(t, y_N) = 0, \qquad U(0, x) = \delta(x - x_0). \nonumber
\end{align}
First assume that $x_0$ lies strictly inside layer $j$, so that $y_{j-1}<x_0<y_j$. A source located exactly at an interface is obtained as the limit of sources inside an adjacent layer. We seek $U(t,x)$ in the form
\vspace{-1em}
\begin{equation}\label{eq:ansatz}
U(t, x) = \ind_{\{y_{j - 1} < x < y_j\}} H_j(t, x, x_0) +  \sum_{i = 1}^{N} \ind_{\{y_{i - 1} < x < y_i\}} U_i(t, x),
\end{equation}
where $H_j$ carries the source and each $U_i$ is source-free with prescribed boundary data,
\begin{alignat}{3} \label{eq:Ui}
\sigma_i^2 \partial_{xx} U_i &= \partial_t U_i,   &\qquad U_i(t, y_{i - 1}) &= f_{i - 1}(t), &\qquad U_i(t, y_i) &= f_i(t), \qquad U_i(0, x) = 0, \\
\sigma_j^2 \partial_{xx} H_j &= \partial_t H_j, &\qquad H_j(t, y_{j - 1}) &= H_j(t, y_j) = 0, &\qquad H_j(0, x) &= \delta(x - x_0), \nonumber
\end{alignat}
with the following set of matching conditions
\begin{align} \label{eq:MatchingCondsContinuity}
U_1(t, y_0) = 0, \qquad U_i(t, y_i) &= U_{i + 1}(t, y_i), \qquad U_N(t, y_N) = 0, \quad i = 1..N - 1, \\
\sigma^2_i \partial_x U_i(t, y_i) &= \sigma^2_{i + 1} \partial_x U_{i + 1}(t, y_i), \quad \quad i \neq j - 1, j, \nonumber \\
\sigma^2_{j - 1} \partial_x U_{j - 1}(t, y_{j - 1}) &= \sigma^2_{j} \partial_x U_{j}(t, y_{j - 1}) +  \sigma^2_{j} \partial_x H_{j}(t, y_{j - 1}), \nonumber \\
\sigma^2_{j} \partial_x U_{j}(t, y_{j}) + \sigma^2_j \partial_x H_j(t, y_j) &= \sigma^2_{j + 1} \partial_x U_{j + 1}(t, y_j), \nonumber
\end{align}
with $f_0 \equiv f_N \equiv 0$ enforcing the Dirichlet barriers. The unknowns are the interface densities $f_1, \dots, f_{N - 1}$.

\subsection{The GIT representation and boundary kernels}

We use the representation obtained by the GIT method in \cite{ItkinMuraveyDBFMF} for the heat equation on a time-dependent interval. Specialized to layer $i$, with the interface densities $f_{i - 1}$ and $f_i$ as boundary values and a zero initial condition, it expresses the gradients of $U_i$ at the ends of the layer through $f_{i - 1}$ and $f_i$. The derivation is given in \cref{app:potentials}.

Introducing the even and odd image kernels
\begin{equation} \label{eq:kernels}
\etaev{i}(t) = \frac{1}{\sigma_i \sqrt{\pi t}} \sum_{n \in \mathbb{Z}} e^{- \frac{(2 n \ell_i)^2}{4 \sigma_i^2 t}}, \qquad \etaod{i}(t) = \frac{1}{\sigma_i \sqrt{\pi t}} \sum_{n \in \mathbb{Z}} e^{- \frac{((2 n + 1) \ell_i)^2}{4 \sigma_i^2 t}},
\end{equation}
the layer gradients take the convolution form
\begin{align} \label{eq:gradients}
\left. \partial_x U_i(t, x) \right|_{x = y_{i - 1}} &= - \int_0^t \left[ \etaev{i}(t - s) \, d f_{i - 1}(s) - \etaod{i}(t - s) \, d f_i(s) \right] = -\etaev{i} * f'_{i - 1} + \etaod{i} * f'_i  \\
\left. \partial_x U_i(t, x) \right|_{x = y_i} &= - \int_0^t \left[ \etaod{i}(t - s) \, d f_{i - 1}(s) - \etaev{i}(t - s) \, d f_i(s) \right] =  -\etaod{i} * f'_{i - 1} + \etaev{i} * f'_i. \nonumber
\end{align}

As shown in \cite{ItkinMuraveyDBFMF}, the kernels \eqref{eq:kernels} are theta functions of the layer SLW. With $\mathsf{q}_i := e^{-\omega_i^2/t}$, $\vartheta_3(\mathsf{q}) := \sum_{n \in \Z} \mathsf{q}^{n^2}$ and $\vartheta_2(\mathsf{q}) := \sum_{n \in \Z} \mathsf{q}^{(n + 1/2)^2}$, they read $\etaev{i}(t) = \vartheta_3(\mathsf{q}_i)/(\sigma_i \sqrt{\pi t})$ and $\etaod{i}(t) = \vartheta_2(\mathsf{q}_i)/(\sigma_i \sqrt{\pi t})$. The image sums converge fast for small $t$. Poisson summation turns them into Fourier series in $e^{-\pi^2 m^2 t/\omega_i^2}$, which converge fast for large $t$.

The Dirichlet kernel on layer $j$ is the classical image sum
\begin{equation}\label{eq:Hjexpl}
H_j(t, x, x_0) = \frac{1}{2 \sigma_j \sqrt{\pi t}} \sum_{n \in \Z} \left[ e^{-\frac{(2n \ell_j + x - x_0)^2}{4 \sigma_j^2 t}} - e^{-\frac{(2n \ell_j + x + x_0 - 2y_{j - 1})^2}{4 \sigma_j^2 t}} \right],
\end{equation}
with boundary gradients
\begin{align}\label{eq:upsilon}
\upsilon^-(t\,|\,x_0) &:= \partial_x H_j \big|_{x = y_{j - 1}} = \frac{1}{2 \sigma_j^3 \sqrt{\pi t^3}} \sum_{n \in \Z} (a + 2n \ell_j)\, e^{-\frac{(a + 2n \ell_j)^2}{4 \sigma_j^2 t}}, \qquad a := x_0 - y_{j - 1}, \\
\upsilon^+(t\,|\,x_0) &:= \partial_x H_j \big|_{x = y_j} = -\frac{1}{2 \sigma_j^3 \sqrt{\pi t^3}} \sum_{n \in \Z} \big( (2n + 1) \ell_j - a \big)\,      e^{-\frac{((2n + 1) \ell_j - a)^2}{4 \sigma_j^2 t}}. \nonumber
\end{align}

Substituting \cref{eq:gradients,eq:upsilon} into the flux conditions \eqref{eq:MatchingCondsContinuity} gives a closed convolution system for $f_1, \dots, f_{N - 1}$,
\begin{align} \label{eq:volterra}
-\sigma_i^2\, \etaod{i} * f'_{i - 1} &+ \big( \sigma_i^2 \etaev{i} + \sigma_{i + 1}^2 \etaev{i + 1} \big) * f'_i - \sigma_{i + 1}^2\, \etaod{i + 1} * f'_{i + 1}
\; = \; h_i(t), \\
h_i = 0 \ (i \ne j - 1, \,j), \qquad h_{j - 1} &= \sigma_j^2\, \upsilon^- (t\,|\,x_0), \qquad h_j = -\sigma_j^2\, \upsilon^+(t\,|\,x_0). \nonumber
\end{align}

When $j=1$, the source layer is adjacent to the lower barrier, $f_0 \equiv 0$,
and the row $i=j-1=0$ is absent. Consequently, only the $h_j$ entry remains.
The case $j=N$ is treated analogously at the upper barrier.
Here and below, $\bm{1}_i$ denotes the $i$-th unit vector of $\mathbb{R}^{N-1}$,
for $i=1,\ldots,N-1$. Throughout the formulas below, terms involving
$\sigma_0$, $\sigma_{N+1}$, $\bm{1}_0$, or $\bm{1}_N$ are omitted.

%=======================================================================
\section{Laplace-domain reduction to a tridiagonal system} \label{sec:laplace}

With $\Lap$ the Laplace transform in $t$ with transform variable $\lambda$, and $f_i(0) = 0$, so that $\Lap f_i' = \lambda \Lap f_i$,
\begin{equation}\label{eq:Letas}
\Lap \etaev{i}(\lambda) = \frac{\coth(\omega_i \sqrt{\lambda})}{\sigma_i \sqrt{\lambda}}, \qquad \Lap \etaod{i}(\lambda) = \frac{1}{\sigma_i  \sqrt{\lambda}\, \sinh(\omega_i \sqrt{\lambda})}.
\end{equation}

\begin{proposition}[source transforms]\label{prop:source}
Let
\begin{equation} \label{eq:gammas}
\gamma_1 := \frac{y_j - x_0}{\ell_j}, \qquad \gamma_2 := \frac{x_0 - y_{j - 1}}{\ell_j}, \qquad \gamma_1 + \gamma_2 = 1.
\end{equation}
Then
\begin{equation}\label{eq:Lupsilon}
\sigma_j^2\, \Lap \upsilon^-(\lambda) = \frac{\sinh\! \big( \gamma_1 \omega_j \sqrt{\lambda} \big)} {\sinh\! \big( \omega_j \sqrt{\lambda} \big)},
\qquad -\sigma_j^2\, \Lap \upsilon^+(\lambda) = \frac{\sinh\! \big( \gamma_2 \omega_j \sqrt{\lambda} \big)} {\sinh\! \big( \omega_j \sqrt{\lambda} \big)}.
\end{equation}
\end{proposition}

\begin{proof}
Write $a := x_0 - y_{j - 1} = \gamma_2 \ell_j$ and $\kappa := \sqrt{\lambda}/\sigma_j$. For $z \ne 0$,
\[
  \Lap\! \left[ \tfrac{z}{2 \sigma^3 \sqrt{\pi t^3}} e^{-z^2/(4 \sigma^2 t)} \right]
  = \sigma^{-2} \sgn(z)\, e^{-|z|\sqrt{\lambda}/\sigma}.
\]
Apply this termwise to \eqref{eq:upsilon}. In $\upsilon^-$ the shifts
$a + 2n \ell_j$ are positive for $n \ge 0$ and negative for $n \le -1$, so
\[
\sigma_j^2 \Lap \upsilon^- = \sum_{n \ge 0} e^{-(a + 2n \ell_j) \kappa} - \sum_{n \ge 1} e^{-(2n \ell_j - a) \kappa} = \frac{e^{-a \kappa} - e^{-(2 \ell_j - a) \kappa}}{1 - e^{-2 \ell_j \kappa}} = \frac{\sinh \big( (\ell_j - a) \kappa \big)}{\sinh(\ell_j \kappa)}.
\]
Since $\ell_j \kappa = \omega_j \sqrt \lambda$ and $(\ell_j - a) \kappa = \gamma_1 \omega_j \sqrt \lambda$,
this is the first identity. In $\upsilon^+$ the shifts $(2n + 1) \ell_j - a$ are
positive for $n \ge 0$ and negative for $n \le -1$, and the same resummation gives
\[
-\sigma_j^2 \Lap \upsilon^+ = \frac{e^{-(\ell_j - a) \kappa} - e^{-(\ell_j + a) \kappa}}{1 - e^{-2 \ell_j \kappa}} = \frac{\sinh(a \kappa)}{\sinh(\ell_j \kappa)},
\]
which is the second.
\end{proof}

\subsection{The tridiagonal interface matrix}

Fix a source point $x_0$ strictly inside layer $j$, where $j \in \{1,\ldots,N\}$ and $y_{j-1} < x_0 < y_j$. The source-layer index $j$ remains fixed throughout the calculation. Taking the Laplace transform of \eqref{eq:volterra} and dividing each equation by $\sqrt{\lambda}$ gives a system of $N-1$ linear equations for the transformed interface densities $g_i = (\Lap f_i)(\lambda)$, $i = 1,\ldots,N-1$:
\begin{equation}\label{eq:Msystem}
\Mmat(\lambda)\, \gvec = \frac{1}{\sqrt{\lambda}} \left[
\frac{\sinh(\gamma_1 \omega_j \sqrt{\lambda})}{\sinh(\omega_j \sqrt{\lambda})}\, \bm{1}_{j - 1} + \frac{\sinh(\gamma_2 \omega_j \sqrt{\lambda})}{\sinh(\omega_j \sqrt{\lambda})}\, \bm{1}_{j} \right], \qquad \gvec = \big( \Lap f_1, \dots, \Lap f_{N - 1} \big)^{\!\top}.
\end{equation}
Here $\bm{1}_k$ denotes the $k$th coordinate vector of $\mathbb{R}^{N-1}$. The right-hand side is supported only at the interior interfaces bordering the source layer. The term containing $\bm{1}_0$ is omitted when $j = 1$, and the term containing $\bm{1}_N$ is omitted when $j = N$. The matrix $\Mmat(\lambda)$ has dimension $(N-1) \times (N-1)$ and is independent of the source location.

The entries of the symmetric tridiagonal matrix \(\Mmat\) are
\begin{equation}\label{eq:Mcoef}
\Mmat_{ii} = D_i = \sigma_i \coth(\omega_i \sqrt{\lambda}) + \sigma_{i + 1} \coth(\omega_{i + 1} \sqrt{\lambda}), \qquad \Mmat_{i, i + 1} = \Mmat_{i + 1, i} = -\sigma_{i + 1} \left/ \sinh(\omega_{i + 1} \sqrt{\lambda}). \right.
\end{equation}
$\Mmat$ is symmetric because the generator is formally self-adjoint on $L^2(d x)$, which is exactly the content of \cref{rem:whatisX}. This is worth stating: the symmetry is not an accident of the algebra.

\begin{myremark}[flux form] \label{rem:fluxform}
Multiplying \eqref{eq:Msystem} by $\sqrt{\lambda}$ gives $\Kmat(\lambda) \gvec = \bm{s}(\lambda)$ with $\Kmat := \sqrt{\lambda}\,\Mmat$. Written with a layer-dependent symbol $\nu_i$,
\begin{equation}\label{eq:Kflux}
\begin{gathered}
\Kmat_{ii} = \sigma_i \nu_i \coth(\omega_i \nu_i) + \sigma_{i + 1} \nu_{i + 1} \coth(\omega_{i + 1} \nu_{i + 1}), \qquad \Kmat_{i, i + 1} = \Kmat_{i + 1, i} = -\frac{\sigma_{i + 1} \nu_{i + 1}}{\sinh(\omega_{i + 1} \nu_{i + 1})}, \\
\bm{s} = \frac{\sinh(\gamma_1 \omega_j \nu_j)}{\sinh(\omega_j \nu_j)}\,\bm{1}_{j - 1}
+ \frac{\sinh(\gamma_2 \omega_j \nu_j)}{\sinh(\omega_j \nu_j)}\,\bm{1}_{j},
\end{gathered}
\end{equation}
with $\nu_i = \sqrt{\lambda}$ for every $i$. Row $i$ is the transformed flux balance at $y_i$. In computations we evaluate $\coth z = (1 + e^{-2z})/(1 - e^{-2z})$ and $1/\sinh z = 2e^{-z}/(1 - e^{-2z})$ with $\Re z \ge 0$, which avoid overflow for large positive $\Re z$. Small arguments require series or an \texttt{expm1}-based evaluation, and poles must be treated by limits or avoided during numerical evaluation.
\end{myremark}

For a source at an interior interface $x_0=y_i$, continuity of the resolvent in the source location gives \eqref{eq:Kflux} with $\bm s=\bm 1_i$. This follows by taking $x_0\to y_i$ from either neighboring layer. The source-layer Dirichlet term vanishes at every fixed interior evaluation point and the source is carried by the interface flux jump. In the time domain, the corresponding interface density generally behaves as $t^{-1/2}$ near zero. Thus the assumption $f_i(0)=0$ used for an interior source is not applied to this case. Interface-started prices and occupation transforms are defined by this resolvent limit.

\begin{lemma}[energy identity]\label{lem:energy}
Let $\lambda \in \mathbb{R}$ lie outside the Dirichlet spectrum of each layer,
so that $\lambda \ne -(m\pi/\omega_i)^2$ for all $m \ge 1$ and $i=1,\ldots,N$.
For $\bm v \in \mathbb{R}^{N-1}$, let $w$ denote the continuous function
satisfying $w(y_i)=v_i$ for $i=1,\ldots,N-1$, $w(y_0)=w(y_N)=0$,
and $\sigma_i^2 w''=\lambda w$ in the interior of each layer. Then

\begin{equation}\label{eq:energy}
\bm v^{\top} \Kmat(\lambda)\,\bm v = \int_{y_0}^{y_N} \Big( \sigma^2(x)\,w'(x)^2 + \lambda\,w(x)^2 \Big)\,d x.
\end{equation}
In particular $\Kmat(\lambda)$ is positive definite for $\lambda > 0$.
\end{lemma}

\begin{proof}
On layer $i$, integration by parts gives
\[
\int_{y_{i - 1}}^{y_i}(\sigma_i^2 w'^2 + \lambda w^2)\,d x = \big[ \sigma_i^2 w w' \big]_{y_{i - 1}}^{y_i}.
\]
The layer solution with end values $v_{i - 1}, v_i$ has the boundary fluxes
\begin{align*}
\sigma_i^2 w'(y_i -) &= \sigma_i \nu_i \big( v_i \coth(\omega_i \nu_i) - v_{i - 1}/\sinh(\omega_i \nu_i) \big), \\
\sigma_i^2 w'(y_{i - 1} +) &= \sigma_i \nu_i \big( v_i/\sinh(\omega_i \nu_i) - v_{i - 1} \coth(\omega_i \nu_i) \big).
\end{align*}
Summing the boundary terms over the layers with $v_0 = v_N = 0$ gives
\eqref{eq:energy}. For $\lambda = -\xi^2 < 0$ read $\nu_i = \mathrm{i} \xi$, so that
$\sigma_i \nu_i \coth(\omega_i \nu_i) = \sigma_i \xi \cot(\omega_i \xi)$ and
$\sigma_i \nu_i/\sinh(\omega_i \nu_i) = \sigma_i \xi/\sin(\omega_i \xi)$ are real.
\end{proof}

%=======================================================================
\section{Spectral inversion I: equal SLW} \label{sec:equal}

Assume throughout this section
\begin{equation}\label{eq:equaltransit}
\omega_1 = \omega_2 = \dots = \omega_N = : \omega.
\end{equation}
Dividing row $i$ of \eqref{eq:Msystem} by
$(\sigma_i + \sigma_{i + 1})/\sinh(\omega \sqrt{\lambda})$ yields
\begin{align} \label{eq:Asystem}
\Amat\! \big( \cosh(\omega \sqrt{\lambda}) \big)\, \gvec &= \frac{1}{\sqrt{\lambda}}
\left[ \frac{\sinh(\gamma_1 \omega \sqrt{\lambda})}{\sigma_{j - 1} + \sigma_j}\, \bm{1}_{j - 1} + \frac{\sinh(\gamma_2 \omega \sqrt{\lambda})}{\sigma_{j} + \sigma_{j + 1}}\, \bm{1}_{j} \right], \\
\Amat(z) = z\,\Imat - \Bmat, \qquad \Bmat_{i, i - 1} &= \frac{\sigma_i}{\sigma_i + \sigma_{i + 1}}, \qquad \Bmat_{i, i + 1} = \frac{\sigma_{i + 1}}{\sigma_i + \sigma_{i + 1}}, \qquad \Bmat_{ii} = 0. \nonumber
\end{align}

The factor $\sinh(\omega \sqrt{\lambda})$ produced by the row normalization cancels the $\sinh(\omega_j \sqrt{\lambda})$ in the denominator of \eqref{eq:Msystem}. Hence the right-hand side of \eqref{eq:Asystem} has no poles, and $\gvec$ is singular only where $\cosh(\omega \sqrt{\lambda}) \in \operatorname{spec}(\Bmat)$.

\subsection{$\Bmat$ as a matrix of hitting probabilities}

\begin{proposition}[probabilistic meaning of $\Bmat$]\label{prop:chain}
The scale density of $X$ is $s'(x) = 1/\sigma^2(x)$, so for the diffusion
started at an interface $y_i$,
\begin{equation}\label{eq:hit}
\Prb \big( \text{hit } y_{i + 1} \text{ before } y_{i - 1} \big) = \frac{\ell_i / \sigma_i^2}{\ell_i/\sigma_i^2 + \ell_{i + 1}/\sigma_{i + 1}^2} = \frac{\omega_i/\sigma_i}{\omega_i/\sigma_i + \omega_{i + 1}/\sigma_{i + 1}}
\;\xrightarrow[\ \omega_i \equiv \omega\ ]{}\; \frac{\sigma_{i + 1}}{\sigma_i + \sigma_{i + 1}} = \Bmat_{i, i + 1}.
\end{equation}
Hence under \eqref{eq:equaltransit} the sequence of interfaces visited by $X$, each different from the previous one, is a Markov chain with transition matrix $\Bmat$, stopped when it reaches $y_0$ or $y_N$. The matrix $\Bmat$ is substochastic: interior rows sum to $1$, rows $1$ and $N - 1$ sum to less than $1$, the deficit being the killing probability at the adjacent barrier.
\end{proposition}

\begin{proof}
Let $s(x) := \int_{y_0}^{x} d \xi/\sigma^2(\xi)$. By \eqref{eq:scale} and \eqref{eq:scaledpars}, $s'(\Phi(y))\,\Phi'(y) = \vs^2(y) \cdot \sqrt 2/\vs^2(y) = \sqrt 2$. Hence $s(\Phi(y)) = \sqrt 2\,(y - Y_0)$ and $s(X_t) = \sqrt 2\,(S_t - Y_0)$. Since $S$ is a martingale, so is $s(X)$, which is the defining property of a scale function.

Fix $i \in \{1, \dots, N - 1\}$, start $X$ at $y_i$, and let $\rho$ be its first exit time from $(y_{i - 1}, y_{i + 1})$. Up to $\rho$ the process $S$ is a Brownian motion run at the clock $\langle S \rangle_t \ge \vs_{\min}^2 t$, where $\vs_{\min} := \min_k \vs_k > 0$, by the Dambis--Dubins--Schwarz theorem. A Brownian motion leaves a bounded interval in finite time almost surely, so $\rho < \infty$ almost surely. The stopped process $s(X_{t \wedge \rho})$ is a bounded martingale. With $p_i := \Prb(X_\rho = y_{i + 1})$, optional stopping gives
\[
s(y_i) = p_i\, s(y_{i + 1}) + (1 - p_i)\, s(y_{i - 1}), \qquad
p_i = \frac{s(y_i) - s(y_{i - 1})}{s(y_{i + 1}) - s(y_{i - 1})} = \frac{\ell_i/\sigma_i^2}{\ell_i/\sigma_i^2 + \ell_{i + 1}/\sigma_{i + 1}^2},
\]
because $s$ is linear with slope $1/\sigma_k^2$ on layer $k$. The second expression in \eqref{eq:hit} follows from $\ell_i/\sigma_i^2 = \omega_i/\sigma_i$, and $\omega_i = \omega_{i + 1}$ gives the third.

With $\vs$ bounded and bounded away from zero, \eqref{eq:S} has a weak solution that is unique in law by the Engelbert--Schmidt theorem, \cite{karatzas1991brownian}. Hence $S$ is a time-homogeneous strong Markov process within the bucket, and so is $X = \Phi(S)$. Let $\tau_0 := 0$. While $X_{\tau_n} = y_k$ with $1 \le k \le N - 1$, let $\tau_{n + 1}$ be the first time after $\tau_n$ at which $X$ hits $y_{k - 1}$ or $y_{k + 1}$. By continuity of paths these are the only grid points that $X$ can reach next, and $\tau_{n + 1} < \infty$ almost surely by the argument above. The chain $Z_n := X_{\tau_n}$ is killed when it reaches $y_0$ or $y_N$. By the strong Markov property at $\tau_n$, $Z$ is a Markov chain with $\Prb(Z_{n + 1} = y_{k + 1} \mid Z_n = y_k) = p_k$ and $\Prb(Z_{n + 1} = y_{k - 1} \mid Z_n = y_k) = 1 - p_k$. Under \eqref{eq:equaltransit}, $p_k = \Bmat_{k, k + 1}$ and $1 - p_k = \sigma_k/(\sigma_k + \sigma_{k + 1}) = \Bmat_{k, k - 1}$. The transition probabilities of $Z$ between the interior states $y_1, \dots, y_{N - 1}$ are therefore the entries of $\Bmat$. For $2 \le k \le N - 2$ both neighbors of $y_k$ are interior, and row $k$ sums to $1$. Row $1$ lacks the entry $1 - p_1$, the probability of reaching $y_0$ before $y_2$. Row $N - 1$ lacks $p_{N - 1}$, the probability of reaching $y_N$ before $y_{N - 2}$.
\end{proof}

\begin{corollary}[in the price coordinate]
$\Bmat_{i, i + 1} = \vs_i/(\vs_i + \vs_{i + 1}) = L_i/(L_i + L_{i + 1})$, which is the natural-scale hitting probability of the martingale $S$ under \cref{rem:transit}.
\end{corollary}

\begin{proof}
By \eqref{eq:scaledpars}, $\sigma_i = 1/\vs_i$, so $\Bmat_{i, i + 1} = \sigma_{i + 1}/(\sigma_i + \sigma_{i + 1}) = \vs_i/(\vs_i + \vs_{i + 1})$. Under \eqref{eq:equaltransit}, $\vs_i = \sqrt 2\, L_i/\omega$ by \eqref{eq:scaledpars}, which gives the second equality. Finally, $\ell_i/\sigma_i^2 = \sqrt 2\, L_i$, so the first expression in \eqref{eq:hit} equals $L_i/(L_i + L_{i + 1}) = (Y_i - Y_{i - 1})/(Y_{i + 1} - Y_{i - 1})$. This is the probability that the martingale $S$, started at $Y_i$, hits $Y_{i + 1}$ before $Y_{i - 1}$.
\end{proof}

\subsection{The spectrum of $\Bmat$}

\begin{proposition}[reversibility and simple real spectrum]\label{prop:spectrum}
Let $\pi_i := \sigma_i + \sigma_{i + 1}$ and $\Pimat := \diag(\pi_1, \dots, \pi_{N - 1})$.
Then $\Pimat \Bmat$ is symmetric, with off-diagonal entries
$(\Pimat \Bmat)_{i, i + 1} = (\Pimat \Bmat)_{i + 1, i} = \sigma_{i + 1}$. Consequently
$\Pimat^{1/2} \Bmat \Pimat^{-1/2}$ is a symmetric Jacobi matrix with strictly
positive off-diagonals, so $\Bmat$ has real and simple eigenvalues
$\zeta_1, \dots, \zeta_{N - 1}$. Since $\Bmat$ is irreducible, nonnegative and
strictly substochastic in rows $1$ and $N - 1$, its spectral radius is less than $1$, hence $\zeta_i \in (-1, 1), \, \varphi_i := \arccos \zeta_i \in (0, \pi)$.
\end{proposition}

\begin{proof}
For $1 \le i \le N - 2$, \eqref{eq:Asystem} gives $(\Pimat \Bmat)_{i, i + 1} = \pi_i\,\Bmat_{i, i + 1} = \sigma_{i + 1}$ and $(\Pimat \Bmat)_{i + 1, i} = \pi_{i + 1}\,\Bmat_{i + 1, i} = \sigma_{i + 1}$. All other entries of $\Pimat \Bmat$ vanish, so $\Pimat \Bmat$ is symmetric. Hence $\Pimat^{1/2} \Bmat \Pimat^{-1/2} = \Pimat^{-1/2}(\Pimat \Bmat) \Pimat^{-1/2}$ is symmetric and tridiagonal, with zero diagonal and off-diagonal entries $\sigma_{i + 1}/\sqrt{\pi_i \pi_{i + 1}} > 0$. Since $\Bmat$ is similar to this real symmetric matrix, its eigenvalues are real and $\Bmat$ is diagonalizable.

Let $\Bmat \bm v = \zeta \bm v$ with $\bm v \ne 0$. Row $i$ reads $\zeta v_i = \Bmat_{i, i - 1} v_{i - 1} + \Bmat_{i, i + 1} v_{i + 1}$, with $v_0 := 0$, and $\Bmat_{i, i + 1} > 0$. So $v_1$ determines $v_2, \dots, v_{N - 1}$ recursively. Hence $v_1 \ne 0$, and every eigenspace of $\Bmat$ is one-dimensional. A diagonalizable matrix whose eigenspaces are one-dimensional has simple eigenvalues.

For the bound, take $\bm v$ real, which is possible because $\zeta$ is real. Since $\Pimat \Bmat \bm v = \zeta\,\Pimat \bm v$, the inequality $2|v_i v_{i + 1}| \le v_i^2 + v_{i + 1}^2$ gives
\[
|\zeta|\,\bm v^{\top} \Pimat \bm v = \big| \bm v^{\top} \Pimat \Bmat \bm v \big|
= \Big| \sum_{i = 1}^{N - 2} 2 \sigma_{i + 1} v_i v_{i + 1} \Big|
\le \sum_{i = 1}^{N - 2} \sigma_{i + 1} \big( v_i^2 + v_{i + 1}^2 \big)
= \bm v^{\top} \Pimat \bm v - \sigma_1 v_1^2 - \sigma_N v_{N - 1}^2 .
\]
As $\bm v^{\top} \Pimat \bm v > 0$ and $\sigma_1 v_1^2 > 0$, this gives $|\zeta| < 1$, which makes the Perron--Frobenius argument in the statement explicit. Finally, $\varphi_i = \arccos \zeta_i \in (0, \pi)$ because $\zeta_i \in (-1, 1)$.
\end{proof}

\subsection{The eigen-expansion of the density}

The interface densities determine $U$ through the layer kernels. The most convenient form of $U$, and the one we implement, is its eigen-expansion, whose
ingredients come from the same tridiagonal data.

\begin{theorem}[closed-form spectral data under equal SLW] \label{thm:equalmodes}
Assume \eqref{eq:equaltransit}. Let $\bm v_i$ be the eigenvector of $\Bmat$ for
$\zeta_i$, scaled so that $\bm v_i^{\top} \Pimat \bm v_i = 1$, and extended by
$v_{i, 0} = v_{i, N} = 0$. Let $\| \cdot \|$ be the norm of $L^2(d x)$ on $(y_0, y_N)$.
The eigenvalues of $-\mathcal{A}$ on $(y_0, y_N)$ with Dirichlet ends are
$\mu = \xi^2 = \varphi^2/\omega^2$, where $\varphi = \omega \xi$ runs over two families.
\begin{enumerate}[itemsep=10pt]

\item[(a)] Interface modes, $\varphi \in \{\varphi_i + 2 \pi k, \ 2 \pi(k + 1) - \varphi_i : k \ge 0\}$ for $i = 1, \dots, N - 1$. On layer $m$ the eigenfunction is
\begin{equation}\label{eq:ifmode}
\phi(x) = \frac{v_{i, m - 1} \sin\! \big( \varphi\,(y_m - x)/\ell_m \big)               + v_{i, m} \sin\! \big( \varphi\,(x - y_{m - 1})/\ell_m \big)}{\sin \varphi},
\end{equation}
and $\|\phi\|^2 = \omega/2$ for every $k$.

\item[(b)] Interface-blind modes, $\varphi = n \pi$ with $n \ge 1$. On layer $m$ the eigenfunction is $\phi(x) = (-1)^{n(m - 1)} \sigma_m^{-1} \sin \big( n \pi(x - y_{m - 1})/\ell_m \big)$, and $\|\phi\|^2 = \frac{\omega}{2} \sum_{m = 1}^N \sigma_m^{-1}$.

\end{enumerate}

The two families are disjoint, every eigenvalue is simple, and
\begin{equation}\label{eq:Useries}
U(t, x; x_0) = \sum_n e^{-\mu_n t}\,\frac{\phi_n(x)\,\phi_n(x_0)}{\|\phi_n\|^2},
\end{equation}
where $n$ runs over both families.
\end{theorem}

\begin{proof}
On layer $m$ an eigenfunction with eigenvalue $\xi^2$ solves $\sigma_m^2 \phi'' = -\xi^2 \phi$, and $\xi \ell_m/\sigma_m = \xi \omega = : \varphi$. If $\sin \varphi \ne 0$, the layer solution is fixed by its end values, which gives \eqref{eq:ifmode}. Flux continuity at the interfaces is the homogeneous form of \eqref{eq:Asystem} at $\sqrt \lambda = \mathrm{i} \xi$, namely $\Bmat \bm v = \cos \varphi\,\bm v$. Hence $\cos \varphi = \zeta_i$ for some $i$ and $\bm v = \bm v_i$, which is family (a). If $\sin \varphi = 0$, every layer solution vanishes at both ends, and flux continuity at $y_m$ gives $\sigma_{m + 1} c_{m + 1} = (-1)^n \sigma_m c_m$ for the amplitudes $c_m$ of $\sin(n \pi(x - y_{m - 1})/\ell_m)$, which is family (b).

For the norm in (a), use
\[
\int_0^1 \sin^2(\varphi s)\,d s = \frac 12 - \frac{\sin 2 \varphi}{4 \varphi}, \qquad
\int_0^1 \sin(\varphi(1 - s)) \sin(\varphi s)\,d s = \frac 12 \left( \frac{\sin \varphi}{\varphi} - \cos \varphi \right).
\]
With $\ell_m = \omega \sigma_m$ one has
\[
\sum_m \ell_m(v_{m - 1}^2 + v_m^2) = \omega\,\bm v^{\top} \Pimat \bm v, \qquad
2 \sum_m \ell_m v_{m - 1} v_m = \omega\,\bm v^{\top} \Pimat \Bmat \bm v = \omega \cos \varphi\,\bm v^{\top} \Pimat \bm v.
\]
The terms in $\sin 2 \varphi/\varphi$ and $\sin \varphi/\varphi$ cancel, and $\|\phi\|^2 = \frac{\omega}{2} \bm v^{\top} \Pimat \bm v$. The norm in (b) is direct. The families are disjoint because $\varphi_i \in (0, \pi)$. The eigenvalues in (a) are simple by \cref{prop:spectrum}, and those in (b) because the amplitude recursion leaves one free constant. The operator is self-adjoint on $L^2(d x)$ with compact resolvent, so \eqref{eq:Useries} is its spectral resolution.
\end{proof}

\begin{myremark}[Alternative representations and the Poisson summation formula]
Under equal SLW, regrouping the spectral series in \eqref{eq:Useries} and applying the Poisson summation formula yields an image expansion that converges rapidly as $t \to 0$. This construction is analogous to that used for the heat equation on domains with moving boundaries in \cite{ItkinMuraveyDBFMF}. Choosing between the spectral and image representations allows efficient numerical evaluation over a broad range of parameter values. Details of the transformation will be presented elsewhere.
\end{myremark}

\begin{myremark}[two equal layers]
Take $N = 2$, $\sigma_1 = \sigma_2$ and $\ell_1 = \ell_2 = \ell$. Then $\Bmat = [0]$, $\zeta_1 = 0$ and $\varphi_1 = \pi/2$. Family (a) of \cref{thm:equalmodes} gives $\omega \xi \in \{\pi/2, 3 \pi/2, 5 \pi/2, \dots\}$, and the interface-blind modes give $\omega \xi \in \{\pi, 2 \pi, \dots\}$. Together they form $\{m \pi/2\}_{m \ge 1}$, the Dirichlet spectrum of a homogeneous strip of width $2 \ell$.
\end{myremark}

\begin{lemma}[interface-blind modes and the Weyl count] \label{lem:weyl}
Under \eqref{eq:equaltransit} the eigenfunctions with $\omega \xi = m \pi$, $m \ge 1$, vanish at \emph{every} interface $y_1, \dots, y_{N - 1}$. They contribute nothing to the interface densities $f_1, \dots, f_{N - 1}$. They enter $U$ through the poles at $\lambda = -(m \pi/\omega)^2$ of the layer kernels, which every layer carries through the factor $1/\sinh(\omega \sqrt \lambda)$ of its Dirichlet-to-interior map, and which $H_j$ carries on layer $j$. For $\Lambda > 0$, let $\mathcal{N}(\Lambda)$ be the number of eigenvalues of $-\mathcal{A}$ below $\Lambda$. The counting function then closes: the interface modes contribute $(N - 1)\,\omega \sqrt{\Lambda}/\pi + O(1)$ eigenvalues below $\Lambda$, the interface-blind modes contribute $\omega \sqrt{\Lambda}/\pi + O(1)$, and the total $N \omega \sqrt{\Lambda}/\pi$ agrees with Weyl's law
\[
\mathcal{N}(\Lambda) \sim \pi^{-1} \int_{y_0}^{y_N} \sqrt{\Lambda}\,d x/\sigma(x).
\]
\end{lemma}

\begin{proof}
The interface-blind eigenfunctions are family (b) of \cref{thm:equalmodes}. On a layer $i \ne j$ the transformed density is
\[
\left[\Lap f_{i - 1} \sinh(\sqrt \lambda(y_i - x)/\sigma_i) + \Lap f_i \sinh(\sqrt \lambda(x - y_{i - 1})/\sigma_i)\right]/\sinh(\omega \sqrt \lambda).
\]
The transforms $\Lap f_i$ are analytic at $\lambda = -(m \pi/\omega)^2$, because there $\cosh(\omega \sqrt \lambda) = \pm 1$, which is not an eigenvalue of $\Bmat$ by \cref{prop:spectrum}. The poles of the density at these points therefore come from the denominator, on every layer, and on layer $j$ also from $H_j$. For the count, each $i$ contributes the interface modes $\varphi = \varphi_i + 2 \pi k$ and $\varphi = 2 \pi(k + 1) - \varphi_i$ below $\omega \sqrt \Lambda$, about $\omega \sqrt \Lambda/\pi$ of them, and the interface-blind modes $\varphi = n \pi$ number $\omega \sqrt \Lambda/\pi + O(1)$.
\end{proof}

%=======================================================================
\section{Spectral inversion II: general SLW} \label{sec:general}

For unequal SLW the matrix of \cref{rem:fluxform} depends nonlinearly on the spectral parameter. This section shows that the spectrum is simple and locates every eigenvalue exactly. The counting formula is a specialization of the Wittrick--Williams principle, \cite{CorridorWittrickWilliams1971}.

\begin{proposition} \label{prop:SL}
The operator $\mathcal{A} u = \partial_x(\sigma^2(x) \partial_x u)$ on $[y_0, y_N]$, with homogeneous Dirichlet conditions at the ends and the transmission conditions $[u] = [\sigma^2 u_x] = 0$ at each $y_i$, where $[\cdot]$ denotes the jump across $y_i$, is self-adjoint on $L^2(d x)$. Its eigenvalues are real, strictly negative, and \emph{simple}.
\end{proposition}

\begin{proof}
Self-adjointness and negativity follow from the energy form
$\int \sigma^2u_xv_x\,d x$ on $H^1_0(y_0, y_N)$. To prove simplicity, consider the pair
$(u, \sigma^2u_x)$, which is continuous across every interface. For
$\mathcal{A} u = -\xi^2u$ with $\xi > 0$, the pair at $y_m$ follows from the pair
at $y_{m - 1}$ through the transfer matrix
\begin{equation}\label{eq:transfer}
\mathsf{T}_m(\xi) =
\begin{pmatrix}
\cos(\omega_m \xi) & \dfrac{\sin(\omega_m \xi)}{\sigma_m \xi} \\
- \sigma_m \xi \sin(\omega_m \xi) & \cos(\omega_m \xi)
\end{pmatrix},
\qquad \det \mathsf{T}_m = 1.
\end{equation}
The Dirichlet condition at $y_0$ forces the initial pair onto the ray spanned by $(0, 1)$. Every eigenfunction is therefore a multiple of one solution, and $\xi^2$ is an eigenvalue of $-\mathcal{A}$ exactly when $\mathcal D(\xi) := [\mathsf{T}_N(\xi) \cdots \mathsf{T}_1(\xi)]_{12} = 0$.
\end{proof}

For real $\xi > 0$ let $\Kmat(\xi)$ denote the flux matrix \eqref{eq:Kflux} at $\lambda = -\xi^2$, that is, with $\nu_i = \mathrm{i} \xi$:
\begin{equation}\label{eq:Kreal}
\Kmat_{ii}(\xi) = \xi \big( \sigma_i \cot(\omega_i \xi) + \sigma_{i + 1} \cot(\omega_{i + 1} \xi) \big),   \qquad \Kmat_{i, i + 1}(\xi) = \Kmat_{i + 1, i}(\xi) = -\xi\,\sigma_{i + 1} \left/ \sin(\omega_{i + 1} \xi). \right.
\end{equation}
It is real and symmetric wherever it is defined. Write $n_-(\Kmat)$ for its
number of negative eigenvalues.

\begin{theorem}[eigenvalue count]\label{thm:count}
Let $0 < \xi_1 < \xi_2 < \dots$ be the square roots of the eigenvalues of
$-\mathcal{A}$, and let $\#S$ denote the number of elements of a finite set $S$. For every $\xi > 0$ with $\omega_i \xi \notin \pi \Z$ for all $i$,
\begin{equation}\label{eq:count}
\#\{n : \xi_n < \xi\} = \sum_{i = 1}^{N} \left\lfloor \frac{\omega_i \xi}{\pi} \right\rfloor + n_- \big( \Kmat(\xi) \big).
\end{equation}
\end{theorem}
\vspace{-1em}
\begin{proof}
Let $\mu = \xi^2$ and $Q_\mu(u) := \int(\sigma^2u_x^2 - \mu u^2)\,d x$ on
$H^1_0(y_0, y_N)$. By the min-max principle the left side of \eqref{eq:count} is
the largest dimension of a subspace on which $Q_\mu$ is negative definite. Split
$H^1_0 = W \oplus V$, where $V$ collects the functions that vanish at every
interface and $W$ the functions that solve $\sigma_i^2w'' = -\mu w$ on each layer.
The split exists because $\mu$ is not a Dirichlet eigenvalue of any layer. The
two parts are $Q_\mu$-orthogonal: for $w \in W$ and $v \in V$, integration by parts
on each layer leaves only boundary terms, and these vanish because $v$ vanishes
at the interfaces. On $V$ the form decouples into the layers, and layer $i$
contributes its Dirichlet eigenvalues $(m \pi/\omega_i)^2$ below $\mu$, of which
there are $\lfloor \omega_i \xi/\pi \rfloor$. On $W$, \cref{lem:energy} at
$\lambda = -\mu$ gives $Q_\mu(w) = \bm v^{\top} \Kmat(\xi) \bm v$, where $\bm v$
collects the interface values of $w$. The negative index of $Q_\mu$ is the sum of
the two negative indices.
\end{proof}

The right side of \eqref{eq:count} costs $O(N)$ operations. The floors are explicit, and by Sylvester's law of inertia $n_-(\Kmat)$ is the number of negative eigenvalues of the diagonal blocks in a pivoted $LDL^{\top}$ factorization of the tridiagonal matrix. Scalar pivots suffice when none vanishes. Zero or unresolved pivots require the safeguards in \cref{app:eigen}. The count is exact for commensurable and incommensurable SLW alike, and bisection at admissible trial points brackets every $\xi_n$ by index. Floating-point evaluation near layer poles requires moving the trial point or higher precision, as discussed in \cref{app:eigen}.

\begin{corollary}[general eigen-expansion]\label{cor:general}
Let $\xi_n$ be located by \cref{thm:count}, and let $\phi_n$ solve
$\mathcal{A} \phi = -\xi_n^2 \phi$ with $(\phi, \sigma^2 \phi_x)(y_0) = (0, 1)$,
propagated by \eqref{eq:transfer}. On layer $m$,
$\phi_n(x) = a_{nm} \cos(k_{nm}(x - y_{m - 1})) + b_{nm} \sin(k_{nm}(x - y_{m - 1}))$
with $k_{nm} := \xi_n/\sigma_m$, where $(a_{nm}, \sigma_m \xi_nb_{nm})$ is the pair
at $y_{m - 1}$. Then \eqref{eq:Useries} holds with $\mu_n = \xi_n^2$ and
\begin{equation} \label{eq:normgen}
\|\phi_n\|^2 = \sum_{m = 1}^N \bigg[ \frac{a_{nm}^2 + b_{nm}^2}{2}\,\ell_m
+ \frac{a_{nm}^2 - b_{nm}^2}{4k_{nm}} \sin(2k_{nm} \ell_m) + \frac{a_{nm} b_{nm}}{2k_{nm}} \big( 1 - \cos(2k_{nm} \ell_m) \big) \bigg].
\end{equation}
\end{corollary}

\begin{proof}
The count \eqref{eq:count} is exact, and by \cref{prop:SL} it increases by one at each $\xi_n$. Bisection on it therefore isolates every $\xi_n$ and misses none. By \cref{prop:SL}, every eigenfunction for $\xi_n$ is a multiple of the solution whose pair $(u, \sigma^2 u_x)$ at $y_0$ is $(0, 1)$. That solution is $\phi_n$. The transfer matrices \eqref{eq:transfer} keep the pair continuous across every interface, which is the transmission condition, and $\mathcal D(\xi_n) = 0$ gives $\phi_n(y_N) = 0$. On layer $m$, $\phi_n$ solves $\sigma_m^2 \phi'' = -\xi_n^2 \phi$, so it has the stated form. Its value $a_{nm}$ and its flux $\sigma_m^2 \phi_n'(y_{m - 1} +) = \sigma_m \xi_n b_{nm}$ at $y_{m - 1}$ form the pair.

For the norm, write $s := x - y_{m - 1}$ and $k := k_{nm}$. The identities $\cos^2 ks = \frac 12 (1 + \cos 2ks)$, $\sin^2 ks = \frac 12 (1 - \cos 2ks)$ and $2 \sin ks \cos ks = \sin 2ks$ give
\[
  \int_0^{\ell_m} \big( a_{nm} \cos ks + b_{nm} \sin ks \big)^2 \, d s
  = \frac{a_{nm}^2 + b_{nm}^2}{2}\,\ell_m + \frac{a_{nm}^2 - b_{nm}^2}{4k} \sin(2k \ell_m) + \frac{a_{nm} b_{nm}}{2k} \big( 1 - \cos(2k \ell_m) \big),
\]
and summing over the layers gives \eqref{eq:normgen}.

The energy form of \cref{prop:SL} is defined on $H^1_0(y_0, y_N)$, which embeds compactly into $L^2(y_0, y_N)$. Hence $\mathcal{A}$ has compact resolvent, the $\phi_n$ form an orthogonal basis of $L^2(d x)$, and the killed Green's function $U$, which is the kernel of $e^{t \mathcal{A}}$, is \eqref{eq:Useries} with $\mu_n = \xi_n^2$. For $t > 0$ the series converges absolutely and uniformly in $x$ and $x_0$. Indeed, \eqref{eq:count} with $n_-(\Kmat) \le N - 1$ gives $\xi_n \ge \pi(n - N + 1)/\sum_i \omega_i$. In the variables $(u, \sigma_m u_x/\xi_n)$ each layer acts on the pair as a rotation, and each interface rescales the second variable by $\sigma_m/\sigma_{m + 1}$. The amplitudes $(a_{nm}^2 + b_{nm}^2)^{1/2}$ of the layers are therefore comparable uniformly in $n$. Since the first term of \eqref{eq:normgen} dominates for large $\xi_n$, $\sup |\phi_n|/\|\phi_n\|$ is bounded in $n$.
\end{proof}

\begin{myremark}[stability of the forward transfer]\label{rem:transferconditioning}
In the variables $(u, \sigma_mu_x/\xi)$ the matrix $\mathsf{T}_m(\xi)$ is a
rotation by the angle $\omega_m \xi$, and at an interface the second variable is
rescaled by $\sigma_m/\sigma_{m + 1}$. For a fixed finite configuration this gives amplitude bounds independent of the mode index, but the constants depend on the volatility contrasts and the number of interfaces. Oscillatory layer solutions alone do not guarantee a well-conditioned product: if two successive phase angles equal $\pi/2$, then $\mathsf T_2\mathsf T_1=\diag(-\sigma_1/\sigma_2,-\sigma_2/\sigma_1)$. Repeated pairs can amplify one component. Renormalizing the propagated pair while retaining its accumulated scale, or using a phase-based formulation, is therefore appropriate when contrasts or layer counts are large. The numerical tests reported in this paper use moderate contrasts.
\end{myremark}

\begin{myremark}[Reduction to the equal-SLW problem] \label{rem:reductiontoequalslw}
A general SLW problem can be approximated by an equal-SLW problem. Let $i = 1,\ldots,N$ label the original layers, with corresponding parameters $\omega_i$. Subdivide each original layer $i$ into $P_i$ equal sublayers, each of which has the corresponding parameter $\omega_i/P_i$. To obtain approximately equal SLWs across all sublayers, choose the positive integers $P_1,\ldots,P_N$ so that
\[
\frac{\omega_1}{P_1} \approx \frac{\omega_2}{P_2} \approx \cdots \approx \frac{\omega_N}{P_N}.
\]
Thus, each ratio in this expression corresponds to a group of sublayers obtained from one original layer.

Finding suitable integers $P_i$ amounts to a simultaneous Diophantine approximation problem, which can be addressed using the Lenstra--Lenstra--Lov\'asz (LLL) lattice basis reduction algorithm, \cite{LenstraLenstraLovasz1982}. For certain configurations of $\omega_i$, this reduction may be computationally more efficient than solving the eigenvalue equation directly, despite the increased number of layers.
\end{myremark}

%=======================================================================
\section{Barrier and no-touch options}\label{sec:barrier}

The double no-touch pays one at $T$ if neither barrier has been hit before. With $P \equiv 1$ in \eqref{eq:pricing} and the eigen-expansion \eqref{eq:Useries}, whose modes come from \cref{thm:equalmodes} or \cref{cor:general},
\begin{equation}\label{eq:dnt}
V = e^{-rT} \sum_n e^{-\mu_nT}\,\frac{\phi_n(x_0)\,c_n}{\|\phi_n\|^2}, \qquad
c_n := \int_{y_0}^{y_N} P \big( \Phi^{-1}(x) \big)\,\phi_n(x)\,d x.
\end{equation}
The same formula prices knock-out Calls and Puts with strike $K$, $P(S) = (S - K)^+$ or $(K - S)^+$. On layer $m$ the map $\Phi^{-1}$ is affine, $\Phi^{-1}(x) = Y_{m - 1} + \vs_m^2(x - y_{m - 1})/\sqrt 2$, so $P \circ \Phi^{-1}$ is piecewise linear, and $c_n$ is a finite sum of the elementary integrals $\int(\alpha + \beta s) \cos(ks)\,d s$ and $\int(\alpha + \beta s) \sin(ks)\,d s$. The strike adds one breakpoint to the payoff, not to the model. In the Laplace domain the density is a sum of exponentials on each layer and the payoff integrals are equally explicit.

\myparagraph{Rebates and first passage.}

The first-passage densities $p_\pm$ were defined in \eqref{eq:rebate}. From \eqref{eq:Useries},
\[
p_-(t) = \sum_n e^{-\mu_nt} \phi_n(x_0)\,\sigma_1^2 \phi_n'(y_0)/\|\phi_n\|^2,
\]
and similarly at $y_N$. In the Laplace domain, with $\nu_i = \sqrt \lambda$,
\begin{equation}\label{eq:Lfpd}
\begin{aligned}
\Lap p_-(\lambda) &= \frac{\sigma_1 \nu_1\,\Lap f_1}{\sinh(\omega_1 \nu_1)}
+ \delta_{j1}\,\frac{\sinh(\gamma_1 \omega_1 \nu_1)}{\sinh(\omega_1 \nu_1)}, \\
\Lap p_+(\lambda) & = \frac{\sigma_N \nu_N\,\Lap f_{N - 1}}{\sinh(\omega_N \nu_N)}
+ \delta_{jN}\,\frac{\sinh(\gamma_2 \omega_N \nu_N)}{\sinh(\omega_N \nu_N)},
\end{aligned}
\end{equation}
where the second terms are the fluxes of $H_j$ from \cref{prop:source} when the source layer touches a barrier, and $\Lap p_\pm (\lambda) = \EE[e^{-\lambda \tau}; \,X_\tau = y_0 \text{ or } y_N]$.

Since $p_- + p_+$ is the density of $\tau$, the identity $\Lap p_- + \Lap p_+ = 1 - \lambda\,\Lap\,\mathrm{DNT}$ holds, where $\mathrm{DNT}$ is the undiscounted double no-touch as a function of maturity. For the rebate \eqref{eq:rebate}, split $\int_0^T = \int_0^\infty - \int_T^\infty$:
\[
\int_0^Te^{-rt} p_-(t)\,d t = \Lap p_-(r) - \sum_n \frac{e^{-(r + \mu_n) T}}{r + \mu_n}\, \frac{\phi_n(x_0)\,\sigma_1^2 \phi_n'(y_0)}{\|\phi_n\|^2},
\]
which is the solution of one tridiagonal system plus an exponentially convergent series.

\myparagraph{Degenerate limits.}

With $N = 1$ the system \eqref{eq:Kflux} is empty, $U = H_1$, and \eqref{eq:Hjexpl} is the classical image formula. Splitting a homogeneous strip into $N$ equal layers must return the same density, which is an exact test of the whole construction. For the vanilla limit $B_\pm \to \pm \infty$, the bound \eqref{eq:trunc} tells how far out the cut-offs must go. A single barrier is the case where only one cut-off is artificial.

%=======================================================================
\section{Corridor products} \label{sec:corridor}

For occupation-time products, the basic state functional is the occupation vector \eqref{eq:occvec}. For a corridor $C$, let $A_t:=\int_0^{t\wedge\tau}\ind_{\{S_s\in C\}}\,ds$. We distinguish three payment conventions. A continuously paid accrual has value $\EE[\int_0^{T\wedge\tau}e^{-rt}\ind_{\{S_t\in C\}}\,dt]$. An amount retained after a barrier hit and paid at $T$ has value $e^{-rT}\EE[A_T]$. A payment forfeited if a barrier is hit, with payoff function $g_p$ of the accrual fraction $A_T/T$, has value $e^{-rT}\EE[g_p(A_T/T);T<\tau]$.

\begin{assumption}\label{ass:aligned}
The corridor $C = (C_-, C_+)$ is a union of layers, i.e.\ $C_\pm \in \{Y_i\}$.
Write $\mathcal{I}_C \subset \{1, \dots, N\}$ for the corresponding index set.
\end{assumption}
This entails no loss of generality: the corridor edges are simply added to the partition. It costs at most two extra layers.

\begin{proposition}[range accrual and corridor swap]\label{prop:accrual}
Let $G_r(x; x_0) := \int_0^\infty e^{-rt} U(t, x; x_0)\,d t = (\Lap U)(r, x; x_0)$
be the $r$-resolvent density of the killed process. Then
\begin{equation}\label{eq:accrual}
  \EE_{S_*}\! \left[ \int_0^{\tau} e^{-rt}\,\ind_{\{S_t \in C\}}\,d t \right]
  = \int_{\Phi(C)} G_r(x; x_0)\,d x.
\end{equation}
The right-hand side is obtained from a \emph{single} solution of
\eqref{eq:Msystem} at $\lambda = r$. No Laplace inversion is required.
\end{proposition}

\begin{proof}
By \eqref{eq:occinvariance}, $\ind_{\{S_t \in C\}} = \ind_{\{X_t \in \Phi(C)\}}$, and $\tau$ is the exit time of $X$ from $(y_0, y_N)$. The integrand is nonnegative, so Tonelli's theorem and the definition of $U$ as the density of $X_t$ on $\{t < \tau\}$ give
\[
\EE_{S_*}\! \left[ \int_0^{\tau} e^{-rt}\,\ind_{\{S_t \in C\}}\,d t \right]
= \int_0^\infty e^{-rt}\, \Prb_{x_0} \big( X_t \in \Phi(C), \ t < \tau \big)\,d t
= \int_{\Phi(C)} \int_0^\infty e^{-rt}\, U(t, x; x_0)\,d t\,d x,
\]
which is \eqref{eq:accrual}. Both sides are finite for $r > -\xi_1^2$, because $\Prb_{x_0}(t < \tau) = \int U\,d x$ decays like $e^{-\xi_1^2 t}$ by \eqref{eq:Useries}.

For the second claim let $r > 0$, let $\gvec = (g_1, \dots, g_{N - 1})$ solve \eqref{eq:Msystem} at $\lambda = r$, and set $g_0 = g_N := 0$. By the layer decomposition of \cref{sec:potentials}, on layer $m$ the transform $G_r = \Lap U(r, \cdot \,; x_0)$ is the solution of $\sigma_m^2 w'' = r w$ with end values $g_{m - 1}$ and $g_m$, plus $\Lap H_j(r, \cdot \,, x_0)$ on the source layer $m = j$. Since $\int_0^{\ell_m} \sinh(\sqrt r\,s/\sigma_m)\,d s = (\sigma_m/\sqrt r)(\cosh(\omega_m \sqrt r) - 1)$, the first part integrates over the layer to $(\sigma_m/\sqrt r) \tanh(\omega_m \sqrt r/2)\,(g_{m - 1} + g_m)$. The integral of $H_j(t, \cdot \,, x_0)$ over layer $j$ is the probability of not having left the layer by time $t$. Its transform is therefore $(1 - w_j(x_0))/r$, where
\[
w_j(x) := \frac{\sinh \big( \sqrt r (y_j - x)/\sigma_j \big) + \sinh \big( \sqrt r (x - y_{j - 1})/\sigma_j \big)}{\sinh(\omega_j \sqrt r)}
\]
solves $\sigma_j^2 w'' = r w$ with $w_j = 1$ at both ends of the layer, so that $w_j(x_0)$ is the transform of the exit time from the layer. By \cref{ass:aligned}, $\Phi(C)$ is a union of layers, and summing these closed forms over its layers expresses the right side of \eqref{eq:accrual} through $\gvec$ alone. Other values $r > -\xi_1^2$ follow by analytic continuation in $r$.
\end{proof}

For a finite maturity $T$, split $\int_0^T = \int_0^\infty - \int_T^\infty$.
With $c^C_n := \int_{\Phi(C)} \phi_n\,d x$ and \eqref{eq:Useries},
\begin{equation}\label{eq:finiteT}
\EE_{S_*}\! \left[ \int_0^{T \wedge \tau} e^{-rt}\,\ind_{\{S_t \in C\}}\,d t \right]
= \int_{\Phi(C)} G_r(x; x_0)\,d x \quad - \sum_n \frac{e^{-(r + \mu_n) T}}{r + \mu_n}\,\frac{\phi_n(x_0)\,c^C_n}{\|\phi_n\|^2}.
\end{equation}
The first term is the resolvent value of \cref{prop:accrual}, and the series
converges exponentially in $n$. Integrating the eigen-series termwise instead
produces the factor $(1 - e^{-(r + \mu_n) T})/(r + \mu_n)$. That is correct, but it
decays only like $1/\mu_n$, and with $c^C_n = O(1/\xi_n)$ for the discontinuous
indicator the termwise series converges like $n^{-2}$. A corridor swap is a
strip of accruals over reset dates and inherits \eqref{eq:finiteT}. A maturity-paid retained linear accrual is obtained from \eqref{eq:finiteT} with $r=0$ inside the expectation and an overall factor $e^{-rT}$.

\subsection{Corridor variance swaps}

Under the normal model \eqref{eq:S}, $\vs$ is volatility in price units. The following proposition therefore concerns corridor quadratic variation of the price, measured in squared price units. This is a possible contract specification, but it differs from the conventional corridor variance swap on log returns, \cite{CarrLewis2004,CorridorLee2010}.

\begin{proposition}\label{prop:corrvar}
Under \cref{ass:aligned},
\begin{equation}\label{eq:corrvar}
  \int_0^{T \wedge \tau} \vs^2(S_t)\, \ind_{\{S_t \in C\}}\,d t
  = \sum_{i \in \mathcal{I}_C} \vs_i^2\, \Lambda_i(T \wedge \tau),
\end{equation}
a linear functional of the occupation vector. Its expectation follows from
\eqref{eq:accrual} layer by layer.
\end{proposition}

\begin{proof}
By \cref{ass:aligned}, $C$ consists of the open layers $(Y_{i - 1}, Y_i)$ with $i \in \mathcal{I}_C$ and the breakpoints between them, and $\vs = \vs_i$ on $(Y_{i - 1}, Y_i)$. For every breakpoint $Y_k$ the occupation times formula gives $\int_0^t \ind_{\{S_s = Y_k\}}\,d \langle S \rangle_s = 0$, and $d \langle S \rangle_s \ge \vs_{\min}^2\,d s$ with $\vs_{\min} := \min_k \vs_k$. Hence $S$ spends zero Lebesgue time at the breakpoints, and almost surely
\[
\int_0^{T \wedge \tau} \vs^2(S_t)\, \ind_{\{S_t \in C\}}\,d t
= \sum_{i \in \mathcal{I}_C} \vs_i^2 \int_0^{T \wedge \tau} \ind_{\{S_t \in (Y_{i - 1}, Y_i)\}}\,d t .
\]
By \eqref{eq:occinvariance} the last integral is $\Lambda_i(T \wedge \tau)$, which proves \eqref{eq:corrvar}. Taking expectations, $\EE_{S_*}[\Lambda_i(T \wedge \tau)]$ is \eqref{eq:finiteT} with $r = 0$ and $C$ replaced by the layer $(Y_{i - 1}, Y_i)$, and the first term of \eqref{eq:finiteT} is \eqref{eq:accrual} for that layer. With the factor $e^{-rt}$ inside the integral the same argument gives the discounted corridor variance.
\end{proof}

For the standard log-return convention, assume $0<B_-\le C_-<C_+\le B_+$, so that the stopped price remains positive. Define the unannualized corridor log-return variance by
\begin{equation}\label{eq:logcorrvar}
Q_C(T):=\int_0^{T\wedge\tau}\ind_{\{S_t\in C\}}\,d\langle\log S\rangle_t=\int_0^{T\wedge\tau}g_C(S_t)\,dt,\qquad g_C(y):=\frac{\vs^2(y)}{y^2}\ind_C(y).
\end{equation}
This is the continuously monitored log-return quantity used in the conventional corridor variance swap, \cite{CorridorLee2010}. In the normal model, $g_C$ varies within each layer. The following representation prices its expected floating leg without approximating $1/y^2$ by a layer constant.

\begin{proposition}[Standard log-return corridor variance]\label{prop:logcorrvar}
Define
\[
\widetilde g_C(x):=g_C(\Phi^{-1}(x)),\qquad c_n^{\log}:=\int_{y_0}^{y_N}\widetilde g_C(x)\phi_n(x)\,dx.
\]
For $r\ge0$, the expected discounted reward is
\begin{equation}\label{eq:logcorrvarprice}
\EE_{S_*}\!\left[\int_0^{T\wedge\tau}e^{-rt}g_C(S_t)\,dt\right]=\int_{y_0}^{y_N}\widetilde g_C(x)G_r(x;x_0)\,dx-\sum_n\frac{e^{-(r+\mu_n)T}}{r+\mu_n}\frac{\phi_n(x_0)c_n^{\log}}{\|\phi_n\|^2}.
\end{equation}
The resolvent and eigenfunctions are the same as for the other single-bucket products. Only the reward integrals change. For a maturity-paid floating leg that retains accrued variance after a barrier hit, evaluate \eqref{eq:logcorrvarprice} at $r=0$ and multiply by $e^{-rT}$.
\end{proposition}

\begin{proof}
Before the stopping time, It\^o's formula gives $d\langle\log S\rangle_t=\vs^2(S_t)S_t^{-2}\,dt$. Since the strip is positive and bounded, $g_C$ is bounded. Integrating it against the killed density and splitting the time integral into its infinite-horizon resolvent and the tail beyond $T$ gives \eqref{eq:logcorrvarprice}, exactly as in \eqref{eq:finiteT}. Positive Dirichlet eigenvalues make the formula well defined also at $r=0$.
\end{proof}

The layer integrals in \eqref{eq:logcorrvarprice} have piecewise smooth integrands and can be evaluated to a prescribed quadrature tolerance. They preserve the exponentially damped tail representation. As an independent identity for the maturity-paid retained amount, define
\begin{equation}\label{eq:logvariancepotential}
F_C(y):=2\int_{C_-}^{C_+}\frac{(y-K)^+}{K^2}\,dK,\qquad F_C''(y)=\frac{2}{y^2}\ind_C(y).
\end{equation}
The identity for $F_C''$ holds away from the corridor edges. Since $F_C'$ is continuous across these edges, the generalized It\^o formula applied up to $T\wedge\tau$ contains no local-time term and yields
\begin{align} \label{eq:logvarianceidentity}
\EE_{S_*}[Q_C(T)] &= \EE_{S_*}[F_C(S_{T \wedge\tau})] - F_C(S_*) \\
&= \int_{y_0}^{y_N} F_C(\Phi^{-1}(x)) U(T,x;x_0)\,dx + \sum_{ \epsilon \in \{-,+\}} F_C(B_\epsilon) \int_0^T p_\epsilon(t)\,dt-F_C(S_*). \nonumber
\end{align}
Thus the same expected variance can be obtained from a terminal payoff and the two undiscounted exit probabilities.

If all accrued variance is forfeited upon a barrier hit, let $\mathcal P_{s,t}$ denote the undiscounted killed transition operator, also allowing tiled volatility, and set $g_C(t,y)=\vs^2(t,y)y^{-2}\ind_C(y)$ in the normal model. The Markov property gives
\begin{equation}\label{eq:logvarianceforfeit}
\EE_{S_*}[Q_C(T);T<\tau]=\int_0^T\mathcal P_{0,t}\!\left[g_C(t,\cdot)\mathcal P_{t,T}1\right](S_*)\,dt.
\end{equation}
For annualization $1/T$ and unit notional, a swap retaining accrued variance and paying a fixed strike $K_{\rm var}$ at $T$ has value $e^{-rT}(\EE[Q_C(T)]/T-K_{\rm var})$. If both legs are paid only on survival, its value is $e^{-rT}(\EE[Q_C(T);T<\tau]/T-K_{\rm var}\Prb(T<\tau))$. These expressions specify both the variance and payment conventions.

In the lognormal model $dS_t=\vs(S_t)S_t\,dW_t$ of \cref{app:drift}, the log-return variance rate is $\vs_i^2$ on layer $i$. Consequently, $Q_C(T)=\sum_{i\in\mathcal I_C}\vs_i^2\Lambda_i(T)$ for the occupation times of that model, and the layer-weighted construction applies directly to the standard contract.

\begin{myremark}[the mean does not separate the methods]\label{rem:honest}
The \emph{mean} occupation is a resolvent integral, and an FD
scheme computes it at a similar cost. \cref{prop:corrvar} alone does not separate
this method from Crank--Nicolson. A direct FD scheme for the law of the accrual carries the accrual as a second state variable. FD can also be applied after transforming in the accrual variable.
\end{myremark}

\subsection{Layer-dependent killing and the law of the accrual}\label{sec:law}

\begin{proposition}[layer-dependent killing preserves the structure]\label{prop:charged}
Let $\bm{q} = (q_1, \dots, q_N)$ be layer-dependent killing rates with $\Re q_i \ge 0$, and set
$\nu_i := \sqrt{\lambda + q_i}$, principal branch. Define the Green's function transform with killing
\begin{equation}\label{eq:Psi}
  \Psi(\lambda, \bm{q}; x, x_0)
  := \int_0^\infty e^{-\lambda t}\,
     \EE_{x_0}\! \left[
       e^{-\sum_{i = 1}^N q_i \Lambda_i(t)}\, \delta(X_t - x)\, ; \ t < \tau
     \right] d t.
\end{equation}
Then the interface values $\Lap f_i := \Psi(\lambda, \bm q; y_i, x_0)$ solve the
flux-form system \eqref{eq:Kflux} with these $\nu_i$, which is symmetric
tridiagonal. On a layer $m \ne j$,
\begin{equation}\label{eq:chargedlayer}
  \Psi = \frac{\Lap f_{m - 1} \sinh \big( \nu_m(y_m - x)/\sigma_m \big)
             + \Lap f_m \sinh \big( \nu_m(x - y_{m - 1})/\sigma_m \big)}{\sinh(\omega_m \nu_m)},
\end{equation}
and on layer $j$ the transform of the source kernel
$H_j^{(\bm q)}(t, x, x_0) = e^{-q_jt} H_j(t, x, x_0)$ is added. Setting $\bm q = 0$
recovers \eqref{eq:Mcoef} through $\Kmat = \sqrt \lambda\,\Mmat$.
\end{proposition}

\begin{proof}
Killing at rate $q_i$ inside layer $i$ turns the layer equation into
$\sigma_i^2 \partial_{xx} U = \partial_tU + q_iU$. Away from the source its
Laplace transform is $\sigma_i^2 \partial_{xx} \Lap U = (\lambda + q_i) \Lap U$, so the
layer solutions are hyperbolic functions of $\nu_i(x - y_{i - 1})/\sigma_i$, which
gives \eqref{eq:chargedlayer}. The transmission conditions do not involve the
killing and are unchanged. Repeating the flux balance of \cref{rem:fluxform}
with these layer solutions gives \eqref{eq:Kflux} with the new $\nu_i$. The
source layer is a Dirichlet problem with constant killing $q_j$, whose kernel is
$e^{-q_jt} H_j$, and its boundary fluxes are those of \cref{prop:source} with
$\nu_j$ in place of $\sqrt \lambda$. Symmetry holds because the $(i, i + 1)$ and
$(i + 1, i)$ entries both come from layer $i + 1$.
\end{proof}

Note that multiplying the kernels $\etaev{i}$ and $\etaod{i}$ by $e^{-q_i t}$ in the convolutions \eqref{eq:gradients} does not yield the correct extension to nonzero killing. These kernels act on $f'$ after integration by parts in $s$, and the factor $e^{-q_i(t-s)}$ generates an additional term in this step. Omitting this term leads to a system with $(\sigma_i/\nu_i)\coth(\omega_i\nu_i)$ on the diagonal and a right-hand side scaled by $1/\lambda$. The inconsistency is apparent in the case of uniform killing, $q_i \equiv q$, whose effect is pure discounting. The correct system must therefore coincide with the system without killing evaluated at $\lambda+q$, whereas the expression obtained by this modification differs by a factor of $(\lambda+q)/\lambda$.

\begin{corollary}[Occupation on survival and occupation at exit]\label{cor:joint}
Set $q_i=\beta\ind_{\{i\in\mathcal I_C\}}$ with $\Re\beta\ge0$, and let $A_t:=\int_0^{t\wedge\tau}\ind_{\{S_s\in C\}}\,ds$. For $\Re\lambda>0$, define $\mathcal H(\lambda,\beta):=\int_{y_0}^{y_N}\Psi(\lambda,\bm q;x,x_0)\,dx$ and $\mathcal H_C(\lambda,\beta):=\int_{\Phi(C)}\Psi(\lambda,\bm q;x,x_0)\,dx$. Then
\begin{equation}\label{eq:occupationtransforms}
\mathcal H(\lambda,\beta)=\int_0^\infty e^{-\lambda t}\EE_{x_0}[e^{-\beta A_t};t<\tau]\,dt,\qquad \EE_{x_0}[e^{-\lambda\tau-\beta A_\tau}]=1-\lambda \mathcal H(\lambda,\beta)-\beta \mathcal H_C(\lambda,\beta).
\end{equation}
Inverting $\mathcal H$ first in calendar time and then in the accrual variable recovers the subprobability measure $\Prb_{x_0}(A_T\in da,T<\tau)$ used to price maturity payments forfeited at either barrier. The second identity gives the joint transform of exit time and occupation at exit. It is a distinct transform from the survivor measure at a fixed maturity.
\end{corollary}

\begin{proof}
Let $A_t := \int_0^{t \wedge \tau} \ind_{\{S_s \in C\}}\,d s$. With $q_i = \beta\, \ind_{\{i \in \mathcal{I}_C\}}$, \eqref{eq:occinvariance} and the zero time at the breakpoints, shown in the proof of \cref{prop:corrvar}, give $\sum_i q_i \Lambda_i(t) = \beta A_t$. Let $\Re \beta \ge 0$ and $\Re \lambda > -\xi_1^2$. Since $|e^{-\beta A_t}| \le 1$ and $\Prb_{x_0}(t < \tau)$ decays like $e^{-\xi_1^2 t}$, integrating \eqref{eq:Psi} in $x$ and exchanging the integrals gives
\begin{equation*}
  \int_{y_0}^{y_N} \Psi\,d x = \int_0^\infty e^{-\lambda t}\, \EE_{x_0} \big[ e^{-\beta A_t}; \ t < \tau \big]\,d t,
  \qquad
  \int_{\Phi(C)} \Psi\,d x = \int_0^\infty e^{-\lambda t}\, \EE_{x_0} \big[ e^{-\beta A_t} \ind_{\{X_t \in \Phi(C)\}}; \ t < \tau \big]\,d t .
\end{equation*}
The first identity is the transform in $t$ of the transform in the accrual of the law of $A_t$ on $\{t < \tau\}$. For the pair $(\tau, A_\tau)$, the process $Z_t := e^{-\lambda t - \beta A_t}$ satisfies $d Z_t = -(\lambda + \beta \ind_{\{X_t \in \Phi(C)\}}) Z_t\,d t$ for $t < \tau$. Integrating from $0$ to $\tau$ and taking expectations gives
\[
  \EE_{x_0} \big[ e^{-\lambda \tau - \beta A_\tau} \big] = 1 - \lambda \int_{y_0}^{y_N} \Psi\,d x - \beta \int_{\Phi(C)} \Psi\,d x ,
\]
the double transform of the exit time and the accrual at exit. Inverting the first identity in $\lambda$ at $t = T$ gives $\EE_{x_0}[e^{-\beta A_T}; \, T < \tau]$. Since $A_T$ takes values in $[0, T]$, its values on the imaginary axis $\beta = -\mathrm{i} u$ form the characteristic function of the measure $\Prb_{x_0}(A_T \in d a, \, T < \tau)$ and determine it. A corridor option that pays $g_p(A_T/T)$ at $T$ on $\{T < \tau\}$ is worth $e^{-rT} \int_0^T g_p(a/T)\, \Prb_{x_0}(A_T \in d a, \, T < \tau)$.
\end{proof}

If all $q_i \equiv q$ then $\nu_i \equiv \sqrt{\lambda + q}$ and
\cref{prop:charged} collapses to the original system with
$\lambda \mapsto \lambda + q$. Uniform killing is nothing but discounting. The
content of \cref{prop:charged} is entirely in the \emph{non-uniform} case,
which is exactly the corridor. A lognormal local volatility produces another set of non-uniform killing rates.

\begin{myremark}[Dimension of the pricing problem]\label{rem:dimcount}
A direct FD formulation for a nonlinear accrual payoff uses the augmented state $(x,A)$ and a degenerate parabolic equation with diffusion in $x$ and transport in $A$. Transformation in $A$ instead gives a family of one-dimensional problems. The present method evaluates each transformed spatial problem through an $(N-1)\times(N-1)$ tridiagonal interface system, followed by time inversion and scalar inversion in the accrual transform variable. A one-dimensional FD solver can also be applied after the same transformation.
\end{myremark}

\subsection{Inverting in the accrual}\label{sec:invert}

Let $A_T:=\int_0^{T\wedge\tau}\ind_{\{S_s\in C\}}\,ds$. Since $0\le A_T\le T$, its transform $\beta\mapsto\EE[e^{-\beta A_T};T<\tau]$ is entire and is bounded by $e^{|\Re\beta|T}$ in the left half-plane. This growth requires care when choosing a numerical inversion in $\beta$. So does the possibility that $A_T$ equals $0$ or $T$ with positive probability. Bounded support alone does not rule out real-axis or deformed-contour methods. We use Fourier inversion on the imaginary axis. For $u\in\mathbb R$, set $\bm q=-\mathrm{i}u\bm\chi$, where $\chi_i:=\ind_{\{i\in\mathcal I_C\}}$, and let $\mathcal B$ be a Bromwich line. Define
\begin{align} \label{eq:charfn}
\Xi(u) &:= \EE_{x_0} \big[ e^{\mathrm{i} uA_T}; \ T < \tau \big] = \frac{1}{2 \pi \mathrm{i}} \int_{\mathcal B} e^{\lambda T}\,\hat \Psi(\lambda; u)\,d \lambda, \\
\hat \Psi(\lambda; u) &:= \int_{y_0}^{y_N} \Psi(\lambda, -\mathrm{i} u \bm \chi; x, x_0) \,dx.
\end{align}

\begin{lemma}[location of the singularities]\label{lem:strip}
Let $\mu_1 > 0$ be the smallest eigenvalue of $-\mathcal{A}$. For $u > 0$ the
singularities of $\lambda \mapsto \hat \Psi(\lambda; u)$ lie in the half-strip
$\{\Re \lambda \le -\mu_1, \ 0 \le \Im \lambda \le u\}$.
\end{lemma}

\begin{proof}
The singularities are eigenvalues of $\mathcal{A} + \mathrm{i} u\,\ind_{\Phi(C)}$. They
lie in its numerical range
$\{\langle \mathcal{A} \phi, \phi \rangle + \mathrm{i} u\,\langle \ind_{\Phi(C)} \phi, \phi \rangle : \|\phi\| = 1\}$,
whose real part is at most $-\mu_1$ and whose imaginary part lies in $[0, u]$.
\end{proof}

At large $u$, standard unshifted Talbot contours can require a large positive real extent to enclose the off-axis poles, increasing the size of $e^{\lambda T}$ and the cancellation in the quadrature. We instead deform $\mathcal B$ into a rectangle open to the left: the bottom edge $\Im\lambda=-\delta$ runs from $\Re\lambda=-\infty$ to $\bar c$, the right edge $\Re\lambda=\bar c$ runs from $\Im\lambda=-\delta$ to $u+\delta$, and the top edge returns to $-\infty$, with $\bar c,\delta>0$. By \cref{lem:strip}, the deformation crosses no singularity. On this contour $|e^{\lambda T}|\le e^{\bar cT}$, uniformly in $u$. This removes the growth of that exponential factor, although resolvent evaluation and quadrature still contribute numerical error. Gauss--Legendre panels are used on the three edges. The horizontal tails are truncated according to exponential damping, and the right edge requires a number of panels proportional to $uT$ for the chosen panel size.

Once $\Xi$ is available, the law of the accrual fraction $\alpha := A_T/T \in[0, 1]$
follows from a cosine expansion on $[0, 1]$. For a piecewise smooth payoff $g_p$ whose cosine series agrees with $g_p$ at every value that $\alpha$ takes with positive probability, with
coefficients $G_k := 2 \int_0^1 g_p(a) \cos(k \pi a)\,d a$,
\begin{equation}\label{eq:cos}
  \EE \big[ g_p(\alpha); \ T < \tau \big] = \tfrac 12G_0\,\Xi(0) + \sum_{k \ge 1} G_k\,\Re\,\Xi(k \pi/T).
\end{equation}
The truncation error after $K_1$ terms is set by the decay of $G_k$ and of $\Xi(k \pi/T)$. For the distribution function, $g_p = \ind_{[0, a]}$ and $G_k = 2 \sin(k \pi a)/(k \pi)$. For an option on the accrual fraction with strike $\kappa \in (0, 1)$, which we call an accrual Call, $g_p(\alpha) = (\alpha - \kappa)^+$ and $G_k = 2(\cos k \pi - \cos k \pi \kappa)/(k \pi)^2$. The zeroth term uses $\Xi(0) = \Prb(T < \tau)$, the undiscounted double no-touch.

\subsection{Removing the endpoint singularities}\label{sec:endpoint}

The truncation error of \eqref{eq:cos} is governed by the decay of $\Xi(k \pi/T)$,
and that decay is governed by the behavior of the law of $\alpha$ at the ends of
$[0, 1]$. When the spot sits on a corridor edge the density of $\alpha$ has inverse
square-root singularities at both ends, $\Xi(k \pi/T)$ decays like $k^{-1/2}$, and
\eqref{eq:cos} converges only algebraically. The singular part can be removed
before expanding.

Write $\Xi_k := \Xi(k \pi/T) = \EE[e^{\mathrm{i} k \pi \alpha}; \,T < \tau]$. Suppose that
the subprobability measure $\Prb(\alpha\in da,T<\tau)$ has probability masses $\varrho_0 := \Prb(\alpha = 0, \, T < \tau)$ at $0$ and $\varrho_1 := \Prb(\alpha = 1, \, T < \tau)$ at
$1$, and a density with the expansions
\[
\sum_{m \ge 0} \mathsf{a}_m a^{m - 1/2}, \text{ as } a \downarrow 0, \qquad
\sum_{m \ge 0} \mathsf{b}_m(1 - a)^{m - 1/2}, \text{ as } a \uparrow 1.
\]
Watson's lemma at each end gives, as $k \to \infty$,
\begin{equation}\label{eq:psiasym}
\Xi_k \sim \varrho_0 + (-1)^k \varrho_1 + \sum_{m \ge 0} \frac{(2m)!\,\sqrt{\pi}}{4^m\, m!}\,(k \pi)^{-m - 1/2} \Big[ \mathsf{a}_m\,e^{\mathrm{i} \pi(2m + 1)/4} + (-1)^k\,\mathsf{b}_m\,e^{-\mathrm{i} \pi(2m + 1)/4} \Big].
\end{equation}

The masses $\varrho_0$ and $\varrho_1$ can be computed separately from barrier survival probabilities. If $S_*\in C$, then $\varrho_1=\Prb_{S_*}(\tau_C>T)$ and $\varrho_0=0$, where $\tau_C$ is the first exit from $C$. If $S_*$ lies outside the closed corridor, $\varrho_0$ is the survival probability to $T$ in the connected component of $(B_-,B_+)\setminus\overline C$ containing $S_*$, and $\varrho_1=0$. At a corridor edge both masses vanish for the nondegenerate diffusion. Thus the existing barrier formulas give $\varrho_0$ and $\varrho_1$ without an asymptotic fit. We keep the terms with $m < M$ and fit the $2M + 2$ real coefficients by least squares to the computed $\Xi_k$ with $K_1/2 \le k \le K_1$, so the fit needs no new data. We then choose real numbers $\varpi_0, \dots, \varpi_{2M - 1}$ such that
\[
\mathcal{S}_M(a) := \frac{1}{\sqrt{a(1 - a)}} \sum_{l = 0}^{2M - 1} \varpi_l\,a^l
\]
has the fitted coefficients $\mathsf{a}_m$ and $\mathsf{b}_m$, $m < M$, in its own
endpoint expansions, which is a linear system of size $2M$. An alternative is to fix $\varrho_0$ and $\varrho_1$ at their independently computed values and fit only the $2M$ continuous-density coefficients.

The substitution $a = (1 - \cos \varphi)/2$ gives $d a/\sqrt{a(1 - a)} = d \varphi$, so the cosine coefficients of $\mathcal{S}_M$ become integrals of smooth functions over $\varphi \in[0, \pi]$. Payoff integrals are smooth between payoff breakpoints, at which the quadrature is split. Composite Gauss--Legendre quadrature evaluates these integrals to a prescribed tolerance. With
\[
\Xi^M_k := \int_0^1e^{\mathrm{i} k \pi a} \mathcal{S}_M(a)\,d a + \varrho_0 + (-1)^k \varrho_1,
\]
the transform of $\mathcal{S}_M$ and of the masses $\varrho_0$ and $\varrho_1$, the expansion is applied to the remainder only:
 \begin{equation} \label{eq:cossub}
 \EE \big[ g_p(\alpha); \,T < \tau \big] = \int_0^1 g_p\,\mathcal{S}_M\,d a + \varrho_0\,g_p(0) + \varrho_1\,g_p(1) + \tfrac 12G_0 \big( \Xi_0 - \Xi^M_0 \big) + \sum_{k = 1}^{K_1} G_k\,\Re \big( \Xi_k - \Xi^M_k \big)+\mathcal R_{K_1}[g_p].
\end{equation}

Here $\mathcal R_{K_1}[g_p]:=\sum_{k>K_1}G_k\Re(\Xi_k-\Xi^M_k)$ is the Fourier truncation remainder. Numerical errors in the computed transforms and integrals are additional.

If the endpoint coefficients and masses are matched exactly and the density is smooth in the interior, the remainder behaves like $a^{M - 1/2}$ and $(1 - a)^{M - 1/2}$ at the ends, so its cosine coefficients decay like $k^{-M - 1/2}$. For the accrual Call, where $G_k = O(k^{-2})$, the truncation error falls from $O(K_1^{-3/2})$ to $O(K_1^{-M - 3/2})$. The fit and the subtraction take a few milliseconds, which is negligible next to the time inversions that produce $\Xi_k$. Errors in fitted coefficients can leave lower-order endpoint terms, and errors in the fitted masses can leave nondecaying transform components. The fit, transform accuracy and changes under refinement of $K_1$ and $M$ must therefore be considered together.

Starting inside the corridor does not remove the probability mass at full accrual. Indeed, $\varrho_1$ can be large for a wide corridor or short maturity, and it contributes $(-1)^k\varrho_1$ to $\Xi_k$. The continuous part can also retain an endpoint singularity. Whether subtraction is useful therefore depends on $\varrho_0$, $\varrho_1$, the endpoint coefficients and the payoff, rather than on the spot's position alone.

%=======================================================================
\section{Greeks}\label{sec:greeks}

Differentiating \eqref{eq:dnt} in $x_0$ acts only on $\phi_n(x_0)$. In the price coordinate, for $S_*$ inside layer $j$, the Greeks $\Delta := \partial V/\partial S_*$, $\Gamma := \partial^2 V/\partial S_*^2$ and $\Theta := -\partial V/\partial T$ are
\begin{align} \label{eq:delta}
\Delta &= \frac{\sqrt 2}{\vs_j^2}\,e^{-rT} \sum_ne^{-\mu_nT}  \frac{\phi_n'(x_0)\,c_n}{ \|\phi_n\|^2}, \qquad \Theta = rV + e^{-rT} \sum_n \mu_ne^{-\mu_nT} \phi_n(x_0) c_n/\|\phi_n\|^2, \\
\Gamma &= \frac{2}{\vs_j^4}\,e^{-rT} \sum_ne^{-\mu_nT} \frac{\phi_n''(x_0) \,c_n}{\|\phi_n\|^2} = -\frac{2}{\vs_j^2}\,e^{-rT} \sum_n \mu_ne^{-\mu_nT} \frac{\phi_n(x_0)\,c_n}{\|\phi_n\|^2}, \nonumber
\end{align}
because $\Phi'(S_*) = \sqrt 2/\vs_j^2$ and $\sigma_j^2 \phi_n'' = -\mu_n \phi_n$. So, $\frac 12 \vs_j^2 \Gamma = rV - \Theta$, which is the pricing equation itself. Gamma therefore requires no computation beyond theta.

\begin{proposition}[the interface kink]\label{prop:kink}
$V$ and $\Delta$ are continuous in $S_*$. At an interface $Y_i$, if $\Gamma(Y_i-)\ne0$, the one-sided gammas satisfy
\begin{equation}\label{eq:gammajump}
  \frac{\Gamma(Y_i +)}{\Gamma(Y_i -)} = \frac{\vs_i^2}{\vs_{i + 1}^2}.
\end{equation}
The weighted identity $\vs_{i+1}^2\Gamma(Y_i+)=\vs_i^2\Gamma(Y_i-)$ holds in all cases. If either one-sided gamma vanishes, both do.
\end{proposition}

\begin{proof}
Continuity of $V$ and $\Delta$ is the $C^1$ matching of \cref{prop:scale}. On
each side of $Y_i$ the pricing equation gives $\frac 12 \vs^2 \Gamma = rV - \Theta$,
and the right side is continuous.
\end{proof}

When the weighted gamma is nonzero and the adjacent volatilities differ, gamma has a model-implied jump. A central second difference spanning the interface approaches a weighted average of the one-sided limits and generally has a nonvanishing error relative to either limit. One-sided differences on an interface-fitted grid avoid this averaging and can converge to the appropriate one-sided gamma.

\subsection{Layer vegas}

A layer vega is the sensitivity of the price to one volatility $\vs_i$ with the breakpoints $Y_0, \dots, Y_N$ held fixed. This changes $\sigma_i$, $\ell_i$, $\omega_i$ and every $y_m$ with $m \ge i$. In particular, perturbing a single volatility moves the model out of the equal-SLW family even when it starts there, so a perturbation of $\Bmat$ alone does not deliver it. A direct approach works in the price coordinate, where the geometry is fixed.

\begin{proposition}[eigenvalue sensitivities]\label{prop:HF}
\begin{equation}\label{eq:HF}
  \frac{\partial \mu_n}{\partial \vs_i} = \frac{2 \mu_n}{\vs_i}\,p_{n, i},
  \qquad
  p_{n, i} := \frac{1}{\|\phi_n\|^2} \int_{y_{i - 1}}^{y_i} \phi_n(x)^2\,d x.
\end{equation}
The weights satisfy $p_{n, i} \in[0, 1]$ and $\sum_ip_{n, i} = 1$, so
$\sum_i \vs_i\,\partial \mu_n/\partial \vs_i = 2 \mu_n$.
\end{proposition}

\begin{proof}
In the price coordinate the eigenproblem reads $-\phi'' = \mu\,w\,\phi$ on
$(B_-, B_+)$ with Dirichlet ends, weight $w := 2/\vs^2$ and $\phi \in C^1$. The form
domain $H^1_0(B_-, B_+)$ and the energy $\int \phi'^2\,d y$ do not depend on $\vs$.
Only the weight does. The Hellmann--Feynman formula for the Rayleigh quotient
gives
\[
\partial \mu_n = -\mu_n \int(\partial w)\,\phi_n^2\,d y \left/ \int w\,\phi_n^2\,d y, \right.
\]
with $\partial w/\partial \vs_i = -4 \vs_i^{-3}$ on layer $i$ and zero elsewhere. The
change of variables $d x = \sqrt 2\,d y/\vs^2$ gives \eqref{eq:HF}.
\end{proof}

The full vega differentiates every term of \eqref{eq:dnt}:
\begin{equation} \label{eq:vega}
\frac{\partial V}{\partial \vs_i} = e^{-rT} \sum_ne^{-\mu_nT} \bigg[ -T\,\frac{\partial \mu_n}{\partial \vs_i}\, \frac{\phi_n(S_*)\,c_n}{\|\phi_n\|^2} + \frac{\partial}{\partial \vs_i} \Big( \frac{\phi_n(S_*)\,c_n}{\|\phi_n\|^2} \Big) \bigg],
\end{equation}
with $\phi_n$ read in the price coordinate. The second term needs the derivative of the eigenfunction. With the normalization $(\phi, \phi_y)(B_-) = (0, 1)$ the eigenfunction is an explicit product of transfer matrices in $(\mu, \vs)$, and its total derivative is $\partial_{\vs_i} \phi + \partial_\mu \phi\,\partial \mu_n/\partial \vs_i$. Every term of \eqref{eq:vega} carries $e^{-\mu_nT}$, so the series converges exponentially. We evaluate the derivatives of the explicit layer formulas by complex-step differentiation. For a real-valued analytic expression, the complex-step quotient with step $h$ has truncation error $O(h^2)$ and avoids subtracting nearby real values, \cite{Squire1998UCV}. Its accuracy also depends on the root continuation, spectral truncation and floating-point implementation. It is not an exact numerical derivative.

\begin{myremark}[the Duhamel form converges slowly]\label{rem:duhamel}
Duhamel's formula gives the same vega as a double series,
\[
\frac{\partial V}{\partial \vs_i} = -\frac{2}{\vs_i}\,e^{-rT} \sum_{m, n}      \frac{\phi_m(x_0)\,O^{(i)}_{mn}\,\mu_nc_n\,J_{mn}}{\|\phi_m\|^2\,\|\phi_n\|^2},
\]
with
\[
O^{(i)}_{mn} := \int_{y_{i - 1}}^{y_i} \phi_m \phi_n\,d x, \quad J_{mn} := \int_0^Te^{-\mu_ms - \mu_n(T - s)}\,d s.
\]
It is exact, but $\mu_nJ_{mn}$ does not decay in $n$. For a knock-out payoff that does not vanish at a barrier, $c_n = O(1/\xi_n)$ and the double series converges only like $1/n$. The scaling identity $\sum_i \vs_i\,\partial V/\partial \vs_i = 2T\,\partial V/\partial T$ for the undiscounted price still holds to rounding after truncation, because the off-diagonal overlaps cancel in the sum over $i$. This makes the slow convergence easy to miss.
\end{myremark}

%=======================================================================
\section{Joining the time buckets}\label{sec:tiled}

We now extend the time-homogeneous construction to the tiled volatility specification \eqref{eq:vspwc} by composing the pricing operators across successive time buckets. Each bucket has its own spectral representation, and the solution is transferred between the corresponding eigenbases at the bucket boundaries. We first derive the backward recursion for terminal payoffs and then extend it to running rewards.

\subsection{The bucket recursion}

Consider the tiled volatility \eqref{eq:vspwc}. On bucket $k$ the model is the time-homogeneous model analyzed above, with the volatilities $\vs_{1, k}, \dots, \vs_{N, k}$. In the price coordinate its generator is $\mathcal{G}_k := \frac 12 \vs_k^2(y)\,\partial_{yy}$, where $\vs_k(y) := \vs_{i, k}$ on layer $i$, and its speed measure is $\mathfrak m_k(d y) := 2\,d y/\vs_k^2(y)$. Its eigenpairs $(\mu^{(k)}_n, \phi^{(k)}_n)$ come from \cref{thm:equalmodes} or \cref{cor:general}. Read in the price coordinate, $\phi^{(k)}_n$ is a linear combination of $\cos(\kappa s)$ and $\sin(\kappa s)$ on layer $i$, with $s := y - Y_{i - 1}$ and $\kappa := \sqrt{2 \mu^{(k)}_n}/\vs_{i, k}$.

The domain does not change from bucket to bucket. In the price coordinate $S$ is its own natural scale, so the transmission condition at every breakpoint is continuity of the value and of its first derivative, whatever the volatilities. The bucket semigroups therefore compose. We write the price at $t_k$ in the eigenbasis of the bucket that produced it, $V(t_k, y) = \sum_n z^{(k + 1)}_n \phi^{(k + 1)}_n(y)$.

\begin{proposition}[bucket recursion]\label{prop:bucket}
For the payoff $P$ let $c_m := \langle P, \phi^{(\bar K)}_m \rangle_{\mathfrak m_{\bar K}}$, where $\langle \cdot, \cdot \rangle_{\mathfrak m_k}$ and $\| \cdot \|_{\mathfrak m_k}$ are the inner product and the norm of $L^2(\mathfrak m_k)$. For $k = 1, \dots, \bar K - 1$ define the transfer matrix
\begin{equation}\label{eq:gram}
\mathbf{Q}^{(k)}_{mn} := \frac{\langle \phi^{(k + 1)}_n, \phi^{(k)}_m \rangle_{\mathfrak m_k}}{\|\phi^{(k)}_m\|^2_{\mathfrak m_k}}.
\end{equation}
With $\Delta t_k := t_k - t_{k - 1}$ and $E_k := \diag \big( e^{-(r + \mu^{(k)}_m) \Delta t_k} \big)$, the price is
\begin{equation}\label{eq:slabrec}
V(0, S_*) = \sum_m z^{(1)}_m \phi^{(1)}_m(S_*), \qquad \bm z^{(\bar K)} = E_{\bar K} \big( c_m/\|\phi^{(\bar K)}_m\|^2_{\mathfrak m_{\bar K}} \big)_m, \qquad \bm z^{(k)} = E_k \mathbf{Q}^{(k)} \bm z^{(k + 1)}.
\end{equation}
\end{proposition}

\begin{proof}
On bucket $k$ the price solves $\partial_t V + \mathcal{G}_k V - rV = 0$ with the same boundary and transmission conditions as on every other bucket, so $V(t_{k - 1}) = e^{-r \Delta t_k} e^{\Delta t_k \mathcal{G}_k} V(t_k)$. The eigenbasis of bucket $k$ is orthogonal and complete in $L^2(\mathfrak m_k)$. Expanding $V(t_k)$ in it and applying the bucket semigroup mode by mode gives \eqref{eq:slabrec}.
\end{proof}

The entries of $\mathbf{Q}^{(k)}$ are integrals over each layer of products of two trigonometric functions with different wave numbers, weighted by $2/\vs_{i, k}^2$. They are elementary, and we evaluate them by Gauss--Legendre quadrature on each layer. To state the spectral truncation error, let $\mathsf P_k^{M_k}$ be the orthogonal projection onto the first $M_k$ bucket modes in $L^2(\mathfrak m_k)$. Then, for every $f \in L^2(\mathfrak m_k)$,
\begin{equation}\label{eq:bucketerror}
\|(I-\mathsf P_k^{M_k})\,e^{\Delta t_k(\mathcal G_k-r)} f\|_{\mathfrak m_k}\le e^{-r\Delta t_k}e^{-\mu_{M_k+1}^{(k)}\Delta t_k}\|f\|_{\mathfrak m_k}.
\end{equation}
Thus retaining all modes with $\mu_m^{(k)}\Delta t_k\le\log(1/\varepsilon_k)$ gives a local relative operator bound $\varepsilon_k$ in this norm. By Weyl's law, the corresponding mode count is approximately $M_k \approx (\sum_i\omega_{i,k }/\pi)\sqrt{\log(1 /\varepsilon_k)/\Delta t_k}$, where $\omega_{i,k}$ is the SLW of layer $i$ on bucket $k$.

The norms vary between buckets. With $C_k:=\max_i(\vs_{i,k+1}/\vs_{i,k})$, one has $\|f\|_{\mathfrak m_k}\le C_k\|f\|_{\mathfrak m_{k+1}}$. If $V_M$ denotes the truncated recursion with exact transfer integrals and $V$ the full recursion, a telescoping product estimate gives
\begin{equation}\label{eq:bucketglobalerror}
\|V(0,\cdot)-V_M(0,\cdot)\|_{\mathfrak m_1}\le e^{-rT}\left(\prod_{k=1}^{\bar K-1}C_k\right)\left(\sum_{k=1}^{\bar K}\varepsilon_k\right)\|P\|_{\mathfrak m_{\bar K}}.
\end{equation}
Quadrature and root-finding errors must be added to this spectral bound. Pointwise prices and Greeks require the corresponding smoothing or differentiated-series estimates. Re-expansion coefficients can decay only algebraically when the spatial regularity imposed by neighboring generators differs. Positive bucket lengths provide exponential damping, but very short buckets require more modes.

Two limits are worth recording. With identical buckets and a fixed eigenfunction normalization $\mathbf{Q}^{(k)}$ is the identity, and \eqref{eq:slabrec} collapses to \eqref{eq:dnt}. With a common time factor, $\vs_{i, k} = a_k \vs_i$, let $a(s) := a_k$ on bucket $k$. The deterministic clock $\mathcal T(t) := \int_0^t a^2(s)\,d s$ then maps the model onto the time-homogeneous one. Writing $\mathcal P_s$ for its undiscounted killed semigroup, a terminal-payoff price is $V(0,S_*)=e^{-rT}(\mathcal P_{\mathcal T(T)}P)(S_*)$.

The maturity in the transition law changes to $\mathcal T(T)$, while discounting remains over calendar time $T$. Equivalently, a homogeneous price discounted at rate $r$ to $\mathcal T(T)$ must be multiplied by $e^{-r(T-\mathcal T(T))}$. Undiscounted accumulated corridor quadratic variation also transforms directly, since $a^2(t)\,dt=d\mathcal T(t)$. If its increments are discounted continuously, the transformed reward retains $e^{-r\mathcal T^{-1}(s)}$. Calendar-time accrual additionally acquires the factor $a^{-2}(\mathcal T^{-1}(s))$. The simple homogeneous-clock identity therefore does not apply unchanged to these discounted running rewards.

\subsection{Running rewards}

Continuously paid accruals and corridor price-variance rewards have expectations of the form
\[
\Omega(0, S_*) := \EE_{S_*} \big[ \int_0^{T \wedge \tau} e^{-rt} g(t, S_t)\,d t \big],
\]
which we call running rewards, with $g$ equal to a constant $g_{i, k}$ on each tile. The accrual has $g_{i, k} = \ind_{\{i \in \mathcal{I}_C\}}$, and the normal-model corridor price variance has $g_{i, k} = \vs_{i, k}^2 \ind_{\{i \in \mathcal{I}_C\}}$. These are the continuous-payment conventions specified in \cref{sec:corridor}. On bucket $k$ let $\Omega^\infty_k$ solve the stationary problem $(r - \mathcal{G}_k) \Omega^\infty_k = g_{\cdot, k}$ with the barrier and transmission conditions.

For $r>0$, on each layer $\Omega^\infty_k$ is the constant $g_{i, k}/r$ plus two exponentials. Its values at the breakpoints solve the flux-form system \eqref{eq:Kflux} at $\lambda = r$, built with the volatilities of bucket $k$, with the right-hand side $\big( \bar g_i \sigma_i \nu_i \tanh(\omega_i \nu_i/2) + \bar g_{i + 1} \sigma_{i + 1} \nu_{i + 1} \tanh(\omega_{i + 1} \nu_{i + 1}/2) \big)_i$, where $\bar g_i := g_{i, k}/r$ and $\nu_i = \sqrt r$.

At $r=0$, the stationary Dirichlet problem remains well defined. On layer $i$, writing $s=x-y_{i-1}$ and $F_{i-1},F_i$ for the interface values, its solution is
\begin{equation}\label{eq:zeroratereward}
\Omega^\infty_k(x)=F_{i-1}+\frac{F_i-F_{i-1}}{\ell_i}s+\frac{g_{i,k}}{2\sigma_i^2}s(\ell_i-s).
\end{equation}
The limiting interface matrix has diagonal $\sigma_i^2/\ell_i+\sigma_{i+1}^2/\ell_{i+1}$, off-diagonal $-\sigma_{i+1}^2/\ell_{i+1}$, and right-hand side $(g_{i,k}\ell_i+g_{i+1,k}\ell_{i+1})/2$. These expressions provide a nonsingular evaluation without dividing by $r$.

On bucket $k$ the value is $\Omega(t) = \Omega^\infty_k + e^{(t_k - t)(\mathcal{G}_k - r)} \big( \Omega(t_k) - \Omega^\infty_k \big)$. Writing $\Omega(t_k) = \Omega^\infty_{k + 1} + \sum_n z^{(k + 1)}_n \phi^{(k + 1)}_n$, with $\Omega^\infty_{\bar K + 1} := 0$ and no transfer term on the last bucket, gives
\begin{equation}\label{eq:runrec}
\bm z^{(k)} = E_k \big( \mathbf{Q}^{(k)} \bm z^{(k + 1)} + \bm \psi^{(k)} \big), \quad \psi^{(k)}_m := \frac{\langle \Omega^\infty_{k + 1} - \Omega^\infty_k, \phi^{(k)}_m \rangle_{\mathfrak m_k}}{\|\phi^{(k)}_m\|^2_{\mathfrak m_k}}, \quad \Omega(0, S_*) = \Omega^\infty_1(S_*) + \sum_m z^{(1)}_m \phi^{(1)}_m(S_*).
\end{equation}

Subtracting $\Omega^\infty_k$ is the bucket version of \eqref{eq:finiteT}. Without it the source would enter through the factors $(1 - e^{-(r + \mu) \Delta t})/(r + \mu)$, and the series would converge only algebraically.

The same recursion applies to the standard log-return corridor variance in the normal model, with $g_k(y)=\vs_k^2(y)y^{-2}\ind_C(y)$. This reward is not constant within a tile, so the preceding constant-source layer formula is replaced by the bucket resolvent
\begin{equation}\label{eq:logrewardstationary}
\Omega_k^\infty(y)=\int_0^\infty e^{-rs}(\mathcal P_s^{(k)}g_k)(y)\,ds.
\end{equation}
Here $\mathcal P_s^{(k)}$ is the undiscounted killed semigroup with bucket-$k$ coefficients held fixed. Its resolvent is obtained from \eqref{eq:Kflux}, and integration against $g_k$ uses the same layer quadrature as \eqref{eq:logcorrvarprice}. The semigroup identity and transfer recursion \eqref{eq:runrec} remain valid for this bounded reward. Setting $r=0$ in the running reward and applying the overall calendar discount $e^{-rT}$ gives a maturity-paid retained amount. The survival-forfeited convention instead uses \eqref{eq:logvarianceforfeit}.

The bucket recursion also provides the corresponding price sensitivities. Delta, gamma and theta at $t=0$ follow from \eqref{eq:slabrec} as in \cref{sec:greeks}, using the modes of the first bucket. Since these modes are expressed in the price coordinate, no factor $\Phi'(S_*)$ is required. Every quantity in \eqref{eq:slabrec} depends analytically on the tile volatilities when the roots for each bucket are continued as in \eqref{eq:vega}. All tile vegas can therefore be obtained by term-by-term differentiation, which we again evaluate using complex-step differentiation.

Computing the distribution of the accrual under tiled volatility requires bucket propagators for operators with imaginary killing rates. These operators are not self-adjoint, so the counting theorem and the orthogonality properties used above no longer apply. A Galerkin representation in the eigenbases of the bucket operators without killing provides a possible approach, which we leave for future work. Time-dependent breakpoints require a further extension. In particular, when the breakpoints are fixed in the spot price and the interest rate differs from the dividend yield, the resulting moving-interface problem is of the type studied in \cite{ItkinLiptonMuravey2022_PhysicaA}.

%=======================================================================
\section{Numerical implementation and benchmarks}\label{sec:numerics}

All computations use a Python implementation of the formulas above, run on one core, which accompanies this paper. On a single time bucket three methods are available. The equal-SLW method computes the eigen-decomposition of $\Pimat^{1/2} \Bmat \Pimat^{-1/2}$ with the LAPACK routine \texttt{dstev} and assembles the modes of \cref{thm:equalmodes}. The general method locates the modes by bisection on \eqref{eq:count} and builds them by \eqref{eq:transfer} and \eqref{eq:normgen}. The Laplace-transform method solves \eqref{eq:Kflux} by the Thomas algorithm at the nodes of the optimized Talbot contour with 32 nodes \cite{Weideman2006Optimizing, Trefethen2006Talbot}, integrates the layer solutions against the payoff in closed form, and sums. Across buckets the recursion \eqref{eq:slabrec} uses the general method in every bucket.

The FD reference is a Crank--Nicolson scheme in the price coordinate, with four half-steps of implicit Euler at the start. Its grid has a node on every breakpoint, and at an interface node it uses finite volumes with the speed measure $2\,d y/\vs^2$. This amounts to the harmonic mean of $\vs^2$ weighted by the two neighboring cells. We call it the fitted scheme. A uniform grid that samples $\vs$ at the nodes gives the naive scheme.

The main test case is a tiled surface on the breakpoints $Y = (80, 85, 89.5, 93.5, 97, 100, 103, 106.5, 110.5, \allowbreak 115, 120)$ with twelve monthly buckets over $T = 1$. Its volatilities are $\vs_{i, k} = \vs^{\mathrm{G}}_i(1.2 - 0.4\,\bar t_k) + 2(\bar t_k - \tfrac 12)(i - 5.5)/4.5$, where $\vs^{\mathrm{G}} = (18, 15, 12.5, 11, 10, 9.5, 10, 11, 12.5, 14)$ is a skewed smile and $\bar t_k$ is the midpoint of bucket $k$. The level falls from $1.18$ to $0.82$ times the smile over the year, and the slope of the skew term changes sign.

Three configurations with a single bucket serve as illustrations. The general configuration uses $\vs^{\mathrm{G}}$ at all times. The equal-SLW configuration uses $\vs_i = \frac{10}{3} L_i$, so that $\omega_i \equiv 0.4243$. The two-layer configuration is an oscillating Brownian motion with $\vs = 12$ on $(20, 100)$ and $\vs = 8$ on $(100, 180)$. Unless stated otherwise $S_* = 101$, $r = 0.02$, the Call strike is $100$, and the corridor is $C = (97, 103)$, the union of layers 5 and 6. Times are the best of several wall-clock runs of the same implementation.

\subsection{Volatility piecewise constant in time}

On the tiled surface the Call is worth $2.7457782018$ and the double no-touch $0.8159150345$. In each bucket we keep the modes with $\mu^{(k)}_m \Delta t_k \le \log(1/\varepsilon)$. \cref{tab:tiled} reports the checks.

\begin{table}[ht]
\centering
\caption{Tiled configuration with twelve monthly buckets. Each row is a test of the bucket recursion, with the absolute difference to its reference and the wall-clock time of the spectral computation.}
\label{tab:tiled}
\begin{tabular}{lcr}
\toprule
test & difference & time (s) \\
\midrule
identical buckets against the time-homogeneous price & $4.4 \times 10^{-15}$ & \\
common time factor against the clock change & $5.8 \times 10^{-15}$ & \\
Call, $\log(1/\varepsilon) = 10$, against $\log(1/\varepsilon) = 90$ & $7.5 \times 10^{-9}$ & 0.08 \\
Call, $\log(1/\varepsilon) = 20$, against $\log(1/\varepsilon) = 90$ & $9.3 \times 10^{-15}$ & 0.08 \\
Call against Crank--Nicolson, extrapolated & $3.1 \times 10^{-9}$ & \\
double no-touch against Crank--Nicolson, extrapolated & $3.3 \times 10^{-10}$ & \\
expected accrual against Crank--Nicolson, extrapolated & $2.0 \times 10^{-10}$ & 0.16 \\
corridor price variance against Crank--Nicolson, extrapolated & $8.4 \times 10^{-9}$ & 0.10 \\
delta against central differences & $1.1 \times 10^{-9}$ & \\
120 tile vegas against central differences, four checks & $2.7 \times 10^{-11}$ & 4.10 \\
\bottomrule
\end{tabular}
\end{table}

With $\log(1/\varepsilon)=20$, corresponding to 19 to 27 modes per bucket, the Call agrees with the more highly resolved spectral result to $9.3\times10^{-15}$. This is an internal convergence check. With $\log(1/\varepsilon) = 10$ the error is $7.5 \times 10^{-9}$.

The time-dependent fitted scheme uses the operator of each bucket and $n$ steps in total. It converges at second order to the spectral values, with Call errors of $-5.7 \times 10^{-5}$, $-1.4 \times 10^{-5}$ and $-3.6 \times 10^{-6}$ at $n = 400$, $800$ and $1600$. The last run takes $0.14$ seconds.

The expected accrual in $C$ is $0.3875968500$, and the expected continuously discounted corridor price variance $\EE \big[ \int_0^{T \wedge \tau} e^{-rt} \vs^2(t, S_t) \ind_{\{S_t \in C\}}\,d t \big]$ is $39.69713630$. With identical buckets the recursion \eqref{eq:runrec} reproduces \eqref{eq:finiteT} to $3.5 \times 10^{-14}$. Delta $0.1631869452$ and gamma $-0.0157540053$ agree with central differences to within their truncation error.

The standard log-return corridor variance \eqref{eq:logcorrvar} has the rate $g_k(y) = \vs_k^2(y)\,y^{-2} \ind_C(y)$, which varies within each layer. Its market-standard floating leg retains the accrued variance and is paid at $T$, so it requires $\EE[Q_C(T)]$, that is, \eqref{eq:runrec} with $r = 0$ and the bucket resolvent \eqref{eq:logrewardstationary}. On each layer we write the bucket resolvent as a particular solution plus the two homogeneous solutions. The particular solution follows from the variation-of-parameters integral, evaluated by 48-point Gauss--Legendre quadrature, and the breakpoint values solve the same tridiagonal system as for a constant reward. At $r = 0$ the bucket resolvent is $F_C(B_+)(y - B_-)/(B_+ - B_-) - F_C(y)$ in every bucket, by \eqref{eq:logvariancepotential}, and the computed one agrees with it to $5 \times 10^{-16}$.

On the tiled surface $\EE[Q_C(T)] = 0.003986997479$, a fair strike of $39.87$ variance points, or $6.31$ volatility points. With $\log(1/\varepsilon) = 10$, that is 13 to 19 modes per bucket, the recursion agrees with $\log(1/\varepsilon) = 90$ to $5.3 \times 10^{-14}$ and takes 83 milliseconds. Of these, 58 milliseconds build the bucket bases, which all products on the surface share. \cref{tab:logcvs} compares the fitted and the naive FD schemes with these values.

\begin{table}[!htb]
\centering
\caption{Expected log-return corridor variance $\EE[Q_C(T)] = 0.003986997479$ on the tiled surface, with the accrued variance retained and paid at $T$. Errors are against the spectral value with $\log(1/\varepsilon) = 90$. The FD runs use $n$ space intervals and $n$ time steps. The fitted scheme has a node on every breakpoint, and the naive scheme uses a uniform grid.}
\label{tab:logcvs}
\begin{tabular}{llr S[table-format=-1.1e-1] S[table-format=-1.1e-1]}
\toprule
method & resolution & {time (s)} & {error} & {relative error} \\
\midrule
spectral, $\log(1/\varepsilon) = 10$ & 13--19 modes & 0.083 & 5.3e-14  & 1.3e-11 \\
spectral, $\log(1/\varepsilon) = 20$ & 19--27 modes & 0.100 & -9.5e-18 & 2.4e-15 \\
fitted FD & $n = 200$  & 0.011 & -2.9e-8  & 7.3e-6  \\
fitted FD & $n = 800$  & 0.049 & -2.1e-9  & 5.2e-7  \\
fitted FD & $n = 3200$ & 0.390 & -1.3e-10 & 3.2e-8  \\
fitted FD, extrapolated & $n = 1600, 3200$ & 0.520 & -9.7e-14 & 2.4e-11 \\
naive FD & $n = 400$  & 0.024 & 3.3e-6  & 8.3e-4  \\
naive FD & $n = 1600$ & 0.120 & 8.3e-7  & 2.1e-4  \\
naive FD & $n = 6400$ & 1.500 & 2.1e-7  & 5.2e-5  \\
\bottomrule
\end{tabular}
\end{table}

The fitted scheme converges at second order. Richardson extrapolation of its runs with 1600 and 3200 intervals agrees with the spectral value to $9.7 \times 10^{-14}$. For this product the fitted scheme is competitive. Its relative error is $7.3 \times 10^{-6}$ after 11 milliseconds and $3.2 \times 10^{-8}$ after 0.39 seconds. This is the situation of \cref{rem:honest}. The expected log-return corridor variance is a resolvent integral, and a scheme fitted to the breakpoints, which include the corridor edges, computes it at a similar cost. The naive scheme converges at first order, and its relative error is still $5.2 \times 10^{-5}$ after 1.5 seconds. With the factor $e^{-rt}$ inside the integral the expected reward is $0.003959426516$, and the fitted scheme shows the same second order, with errors within two percent of those in \cref{tab:logcvs}. The forfeit convention \eqref{eq:logvarianceforfeit} is not tested here. There the reward is weighted by the probability of survival to $T$, so it is not a running reward of the form \eqref{eq:runrec}. Its numerical treatment is left for future work.

\subsection{A single bucket: accuracy at the interfaces}

On a single bucket the three methods agree. On the equal-SLW configuration the two spectral methods agree to $4 \times 10^{-15}$ for a double no-touch, a Call and a Put at $T = 0.1$, $1$ and $5$. The Laplace-transform method agrees with both to $6 \times 10^{-13}$, which is the accuracy of the 32-node Talbot rule. The general method reproduces the 40 lowest closed-form eigenvalues to a relative $3.8 \times 10^{-15}$. A homogeneous strip split into $N = 1$, $2$ or $5$ equal layers reproduces the closed-form double no-touch to $3 \times 10^{-13}$.

Weyl's law gives the estimate $M \approx (\sum_i \omega_i/\pi) \sqrt{ \log(1/\varepsilon) /T}$ for the modes retained at damping threshold $\varepsilon$. An absolute price tolerance also depends on the payoff and evaluation point. On the general configuration with $\varepsilon = e^{-32}$ this gives 262 modes at $T = 10^{-3}$ and 8 at $T = 1$, exactly the counts used. The Laplace-transform method costs about two milliseconds per price over the maturities tested. The eigen-series costs 6 to 21 milliseconds, almost all of it in the search for the modes. Once the modes are known, it prices every maturity, strike and Greek at the cost of an inner product.

\cref{tab:cn} compares the fitted and the naive schemes on the general configuration, with $n$ space intervals and $n$ time steps. The fitted scheme converges at second order. The naive scheme converges at first order, and on the Call its error at $n = 1600$ is $1.5 \times 10^{-3}$, about four hundred times the fitted error.

\begin{table}[ht]
\centering
\caption{General configuration. Errors of Crank--Nicolson against the spectral reference
double no-touch $0.819596756784$ and Call $2.807218341507$, and the
time of the fitted scheme for the Call.}
\label{tab:cn}
\begin{tabular}{r S[table-format=-1.1e-1] S[table-format=-1.1e-1]
                  S[table-format=-1.1e-1] S[table-format=-1.1e-1] r}
\toprule
 & \multicolumn{2}{c}{double no-touch} & \multicolumn{2}{c}{Call} & \\
\cmidrule(lr){2-3}\cmidrule(lr){4-5}
$n$ & {naive} & {fitted} & {naive} & {fitted} & time (ms) \\
\hline
100  & 9.1e-4  & -1.6e-5 & 1.2e-4  & -9.7e-4 & 3.4 \\
200  & 1.9e-4  & -4.0e-6 & -5.7e-3 & -2.5e-4 & 7.5 \\
400  & 1.1e-4  & -9.3e-7 & -6.1e-3 & -6.2e-5 & 16.8 \\
800  & 5.4e-5  & -2.3e-7 & -3.1e-3 & -1.6e-5 & 43.8 \\
1600 & 2.7e-5  & -5.8e-8 & -1.5e-3 & -3.9e-6 & 122.9 \\
\bottomrule
\end{tabular}
\end{table}

For the example considered here, gamma jumps at the interface. With the spot on $Y_5 = 100$, where $\vs$ drops from $10$ to $9.5$, the exact one-sided gammas are $-0.0091996$ and $-0.0101935$. Their ratio is $(10/9.5)^2$, as in \eqref{eq:gammajump}, to $5 \times 10^{-13}$. A central second difference on the fitted grid returns their mean, $-0.0096965$, which is wrong by five percent on either side at every resolution. One-sided differences on the fitted grid recover both values to $5 \times 10^{-7}$.

The ten spectral layer vegas of the Call take 20 milliseconds and agree with central differences to $3 \times 10^{-11}$. The Duhamel double series of \cref{rem:duhamel} misses individual vegas by $3.2$ percent. With $r = 0$, Monte Carlo with $10^5$ Euler paths and a Brownian-bridge correction at the barriers gives the double no-touch and the Call within about one standard error of the exact values. The standard errors are $1.2 \times 10^{-3}$ and $1.3 \times 10^{-2}$, and the run takes 27 seconds.

\subsection{The law of the accrual on a single bucket}

The law of the accrual is available when the volatility does not change over the accrual period. The expected discounted accrual in $C$ is $1.105702839$ at $T = \infty$, from the solution of one tridiagonal system. At finite maturity \eqref{eq:finiteT} agrees with a Talbot inversion of the resolvent to $10^{-13}$ with 4 to 18 modes. The termwise-integrated series with 58 modes is off by $3.2 \times 10^{-7}$.

With the killing rate $\beta$ on the corridor layers, the system \eqref{eq:Kflux} gives $\EE[e^{-\beta A_T}; \,T < \tau]$ after a Talbot inversion. A fitted Crank--Nicolson solution of the Feynman--Kac equation $u_\theta = \tfrac 12 \vs^2u_{yy} - \beta \ind_Cu$, extrapolated from 800 and 1600 intervals, agrees with it to $7 \times 10^{-11}$ or better for $\beta = 0.5$, $5$ and $50$.

The two-layer configuration started at the interface, with the corridor $(100, 180)$, has a law in closed form. Far from the barriers the accrual is the occupation time of the upper half-line by an oscillating Brownian motion. With $N = 2$ the system \eqref{eq:Kflux} is scalar and gives Lamperti's double transform with index $\frac 12$. The accrual fraction then has the distribution function $\frac{2}{\pi} \arctan \big( \frac{\vs_+}{\vs_-} \sqrt{a/(1 - a)} \big)$, where $\vs_- = 12$ and $\vs_+ = 8$ are the volatilities below and above the interface. Its mean, $\vs_-/(\vs_- + \vs_+) = 0.6$, shows that the process spends more time where it moves more slowly.

The characteristic function \eqref{eq:charfn} agrees with the closed form to $2.1 \times 10^{-11}$ for $u$ up to $400$. \cref{tab:lamperti} reports the cosine expansion \eqref{eq:cos} with $K_1$ terms, plain and with $M$ endpoint terms removed at each end as in \cref{sec:endpoint}. The plain expansion converges algebraically, because the density has inverse square-root singularities at both ends of $[0, 1]$. With $M = 1$ the error of the accrual Call falls by a factor of $5.7$ per doubling of $K_1$, the predicted $K_1^{-5/2}$.

With $M = 4$, $K_1 = 64$ reaches $2.2 \times 10^{-11}$ for the distribution function and $3.9 \times 10^{-13}$ for the accrual Call in $0.39$ seconds. The plain expansion needs $512$ terms and $18$ seconds to reach $4.7 \times 10^{-5}$. The floor $2.1 \times 10^{-11}$ belongs to the test, since the barriers at $20$ and $180$ remove about $2 \times 10^{-11}$ of mass that Lamperti's law keeps.

\begin{table}[ht]
\centering
\caption{Two-layer configuration started at the interface. Largest absolute error
of the cosine expansion with $K_1$ terms against Lamperti's law, plain and with $M$
endpoint terms removed at each end, for the distribution function at
$a \in \{0.05, 0.25, 0.5, 0.75, 0.95\}$ and for $\EE[(\alpha - \kappa)^+]$ at
$\kappa \in \{0.25, 0.5, 0.75\}$. The last column is the time to compute $\Xi$ at the
$K_1 + 1$ frequencies.}
\label{tab:lamperti}
\begin{tabular}{r S[table-format=1.1e-2] S[table-format=1.1e-2]
                  S[table-format=1.1e-2] S[table-format=1.1e-2] r}
\toprule
$K_1$ & {plain} & {$M = 1$} & {$M = 2$} & {$M = 4$} & time (s) \\
\midrule
\multicolumn{6}{l}{\emph{distribution function}} \\
16  & 1.0e-2  & 1.5e-4  & 3.2e-6   & 6.0e-8   & 0.07 \\
32  & 1.3e-3  & 1.1e-5  & 1.5e-7   & 4.4e-10  & 0.17 \\
64  & 2.1e-3  & 6.6e-6  & 6.2e-8   & 2.2e-11  & 0.39 \\
128 & 2.5e-4  & 4.6e-7  & 2.1e-9   & 2.1e-11  & 1.40 \\
256 & 2.8e-4  & 2.0e-7  & 5.1e-10  & 2.1e-11  & 4.97 \\
512 & 4.7e-5  & 1.4e-8  & 2.9e-11  & 2.1e-11  & 18.00 \\
\midrule
\multicolumn{6}{l}{\emph{accrual Call}} \\
16  & 7.1e-4  & 1.6e-5  & 1.3e-7   & 1.2e-9   & \\
32  & 2.5e-4  & 2.8e-6  & 6.3e-9   & 2.7e-11  & \\
64  & 8.9e-5  & 4.9e-7  & 1.9e-10  & 3.9e-13  & \\
128 & 3.2e-5  & 8.7e-8  & 3.3e-12  & 4.9e-15  & \\
256 & 1.1e-5  & 1.5e-8  & 1.2e-12  & 2.2e-16  & \\
512 & 3.9e-6  & 2.7e-9  & 1.5e-13  & 1.9e-16  & \\
\bottomrule
\end{tabular}
\end{table}

For the general configuration with $r=0$, \cref{tab:law} reports the survivor occupation measure and the accrual Call. Between $K_1=128$ and $K_1=256$, the distribution function changes by at most $2.4\times10^{-7}$ and the accrual Call by $4\times10^{-10}$. The plain expansion is sufficient for these reported evaluation points and tolerances. The endpoint subtraction brings no clear gain in this experiment. The interpretation in \cref{sec:endpoint} allows for a nonzero probability mass at full accrual and does not assume a smooth density at both endpoints.

The competing FD scheme carries the running accrual as a second state variable. It marches backwards on the fitted grid with implicit Euler. Its accrual grid has the spacing of the time step, so the transport in the accrual is an exact shift by one cell for nodes inside the corridor. The scheme is first order in time. With 400 intervals and 1000 steps its error is $9.3 \times 10^{-4}$ in the distribution function and $3.4 \times 10^{-4}$ in the accrual Call, after 46 seconds.

Richardson extrapolation of the last two FD runs agrees with the spectral values to $2.2 \times 10^{-5}$, which is an independent confirmation. The $K_1=32$ calculation takes $0.7$ seconds. Against the $K_1=256$ reference, its largest reported CDF discrepancy is about $3.6\times10^{-5}$ and its accrual-Call discrepancy is about $3.0\times10^{-7}$. The extrapolated FD values have a smaller largest CDF discrepancy, about $2.2\times10^{-5}$, but a larger accrual-Call discrepancy, about $6.9\times10^{-6}$.

\begin{figure}[ht]
\centering
\includegraphics[width=\textwidth]{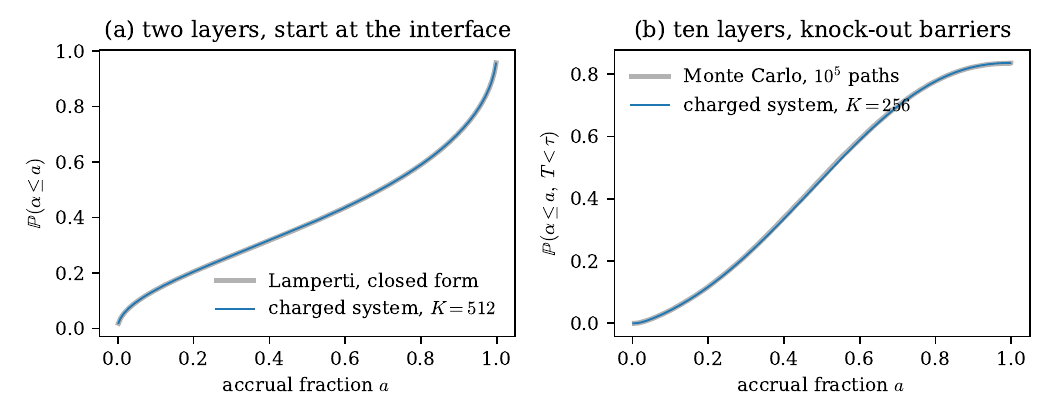}
\caption{Distribution function of the accrual fraction. (a) Two-layer configuration started at the interface, against Lamperti's law. (b) General configuration with knock-out barriers on the event $\{T < \tau\}$, against the empirical distribution of $10^5$ Monte Carlo paths with 4000 steps. Both curves in (b) end at the survival probability $0.836$.}
\label{fig:law}
\end{figure}

Monte Carlo with $10^5$ paths estimates the accrual Call with a low bias. Pooling two seeds at each step size, the bias is $(-5.8 \pm 1.7) \times 10^{-4}$ with 1000 steps per year and $(-1.25 \pm 0.24) \times 10^{-3}$ with 4000. Refining the step did not reduce it over this range, and we have not analyzed its source. \cref{fig:law} compares the distribution functions with Lamperti's law and with the Monte Carlo sample.

\begin{table}[ht]
\centering
\small
\caption{General configuration with knock-out barriers, $r = 0$. Distribution
function of the accrual fraction $\alpha$ on $\{T < \tau\}$, and the accrual Call.
The FD scheme carries the accrual as a second state variable,
with $n$ space intervals and $n_t$ time steps. Monte Carlo (MC) uses $10^5$ paths,
with one standard error in units of the last digit.}
\label{tab:law}
\resizebox{\linewidth}{!}{%
\begin{tabular}{lrccccc}
\toprule
method & time (s) & $a = 0.2$ & $a = 0.4$ & $a = 0.6$ & $a = 0.8$ & $\EE[(\alpha - 0.5)^+]$ \\
\midrule
spectral, $K_1 = 32$  & 0.72 & 0.1169325 & 0.3354596 & 0.5878210 & 0.7767606 & 0.06368258 \\
spectral, $K_1 = 64$  & 1.85 & 0.1169191 & 0.3354984 & 0.5878174 & 0.7767402 & 0.06368231 \\
spectral, $K_1 = 128$ & 5.57 & 0.1169254 & 0.3354954 & 0.5878193 & 0.7767391 & 0.06368228 \\
spectral, $K_1 = 256$ & 17.30 & 0.1169257 & 0.3354954 & 0.5878192 & 0.7767391 & 0.06368228 \\
\midrule
FD, $n = 200$, $n_t = 250$  & 1.18 & 0.1157677 & 0.3330689 & 0.5841291 & 0.7737981 & 0.06504960 \\
FD, $n = 400$, $n_t = 500$  & 9.22 & 0.1163634 & 0.3342896 & 0.5859611 & 0.7752508 & 0.06437300 \\
FD, $n = 400$, $n_t = 1000$ & 46.00 & 0.1166364 & 0.3348871 & 0.5868941 & 0.7760061 & 0.06402420 \\
FD, extrapolated            &      & 0.1169094 & 0.3354846 & 0.5878271 & 0.7767614 & 0.06367540 \\
\midrule
MC, 1000 steps & 7.5  & 0.1168(10) & 0.3356(15) & 0.5880(16) & 0.7761(13) & 0.06332(34) \\
MC, 4000 steps & 27.0 & 0.1175(10) & 0.3391(15) & 0.5900(16) & 0.7772(13) & 0.06256(34) \\
\bottomrule
\end{tabular}}
\end{table}

\subsection{Comparison of methods}

\begin{table}[ht]
\centering
\small
\caption{Where each method wins. General configuration at $T = 1$ unless stated.
Entries give reported absolute discrepancies and wall-clock times. For the accrual law, the FD column uses implicit Euler in the augmented state. The other FD price tests use Crank--Nicolson. The sensitivity row uses bumped FD solves.}
\label{tab:compare}
\resizebox{\linewidth}{!}{%
\begin{tabular}{llll}
\toprule
 & This paper & FD implementation & Monte Carlo \\
\midrule
Call, single maturity       & $\lesssim10^{-12}$ in validation, 2.0 ms             & $3.9 \times 10^{-6}$, 123 ms    & $\pm 1.3 \times 10^{-2}$, 27 s \\
Double no-touch, $T = 5$    & spectral reference, 8.6 ms             & $8.5 \times 10^{-8}$, 228 ms    & not run \\
Gamma at an interface       & analytic series, both sides & one-sided: $5\times10^{-7}$              & not run \\
Ten layer vegas             & $3\times10^{-11}$ vs differences, 20 ms & 20 bumped solves, about 2.5 s        & not run \\
Corridor accrual \emph{law} & $3.6 \times 10^{-5}$, 0.7 s & $9.3 \times 10^{-4}$, 46 s (2D) & observed bias, 27 s \\
Tiled surface, $T = 1$     & $9.3\times10^{-15}$ vs refined series, 0.08 s        & $3.6 \times 10^{-6}$, 140 ms    & not run \\
Accrual law, time-dependent & no                        & yes (2D)                      & yes \\
American exercise           & no                        & yes                           & yes \\
\bottomrule
\end{tabular}}
\end{table}

\cref{tab:compare} collects the comparisons. In the single-vanilla test our Laplace-transform method is faster and more accurate than the fitted Crank--Nicolson implementation, but we do not rely on this comparison. Both are Python implementations, and a compiled FD code would narrow the gap. The claims we defend are four. The method gives exact representations whose numerical evaluation provides a highly resolved reference solution. The eigen-series evaluates $e^{-\mu_nt}$ directly instead of marching, although it needs more modes at shorter maturities. The Laplace-transform method has approximately constant cost over the tested maturities. It gives analytic representations for Greeks, including the one-sided gammas at the interfaces.

Finally, it delivers the occupation law without the extra state dimension of \cref{rem:dimcount}. At a jump of $\sigma$ the flux $\sigma^2 u_x$ is continuous but $u_x$ generally is not, so the tested uniform-grid FD scheme loses order at exactly the points that matter here, while the GIT representation resolves the kinks exactly by construction. FD methods win where the model leaves the present framework, namely continuous time dependence, the law of the accrual under time-dependent volatility, and early exercise. One-sided differentiation on an interface-fitted grid can recover the appropriate gamma limits, and discrete sensitivity equations can replace repeated bumped solves for vegas.

The accrual-law comparison uses a direct FD solver in price and accrued time. Transformation in accrued time is equally available to FD: for each Fourier frequency $u$, one may solve $w_\theta=\tfrac12\vs^2w_{yy}+\mathrm{i}u\ind_Cw$ with $w(0,\cdot)=1$ and absorbing barriers, then perform the same outer Fourier inversion. That one-dimensional complex FD formulation is not benchmarked here. The reported factor of about sixty therefore applies to the direct augmented-state implementation in \cref{tab:law}, rather than to FD methods as a class. The unexplained Monte Carlo discrepancy is likewise specific to the tested implementation. Continuous time dependence, nonlinear accrual payoffs under time-dependent volatility and early exercise remain outside the implementation developed here.

\subsection{Calibration}

We calibrate the ten layer volatilities to ten synthetic vanilla quotes with $r = 0$. The outer breakpoints $80$ and $120$ are replaced by $20$ and $180$, so the barriers act as cut-offs. If the outermost volatilities are continued beyond them, \eqref{eq:trunc} bounds the exit probability by $2\,\mathrm{erfc} \big( 80/(18 \sqrt 2) \big) \approx 1.8 \times 10^{-5}$. For the unbounded Call and Put payoffs, the price-error bound is instead \eqref{eq:truncvanilla}, with coefficient $\max\{P(20),P(180)\}+18$. The quotes are out-of-the-money Puts and Calls at $T = 1$ with $S_* = 100$ and one strike per layer, namely $83$, $87$, $91.5$, $95$, $98.5$, $101.5$, $104.5$, $108.5$, $112.5$ and $117$, generated with the volatilities of the general configuration. Both the quotes and calibrated prices use the same cut-offs, so the calibration residual measures recovery within the truncated model.

Newton's method with the spectral vega Jacobian \eqref{eq:vega}, started from the flat $\vs_i \equiv 12$, reduces the largest pricing error to $0.73$, $0.13$, $5.6 \times 10^{-3}$, $2.6 \times 10^{-5}$, $6.4 \times 10^{-10}$ and $1.7 \times 10^{-14}$ in successive iterations. The recovered volatilities match the generating ones to $6.6 \times 10^{-12}$, and the whole calibration takes $0.9$ seconds.

The observed convergence is consistent with local quadratic convergence when the Jacobian is nonsingular and evaluated sufficiently accurately. The loop closes: the model reprices its calibration instruments to rounding error, using the same kernel to produce the prices and their derivatives. Market quotes are generally not reproduced exactly by a step model, and the same Jacobian then drives a least-squares fit.

%=======================================================================
\section{Conclusion}\label{sec:conclusion}

We priced corridor options and corridor swaps under a local volatility that is piecewise constant in the price and in time. The spatial scale transformation turns the pricing problem into a layered heat equation, and the GIT method reduces it to a symmetric tridiagonal system of fixed size $N - 1$ in the Laplace domain, where $N$ is the number of layers. Under equal SLW, a row normalization gives a fixed matrix of hitting probabilities between neighboring interfaces, and its eigenpairs together with the eigenfunctions that vanish at every interface give the spectrum in closed form.

Layer-dependent killing preserves the tridiagonal structure. This gives transforms of the joint law of exit time and corridor occupation. The single-bucket law of the accrual follows by inverse Laplace transformation in time followed by Fourier inversion in the accrual variable. The same formulation covers the lognormal model through an exponential change of the dependent variable and a particular choice of killing rates. The inversion accounts separately for possible probability masses at zero and full accrual. In the reported tests with the initial price at a corridor edge, accurately subtracting the transform contributions of the density's endpoint singularities improves convergence. Greeks and layer vegas follow by differentiating the spectral representations. Their numerical evaluation uses truncation and, for vegas, complex-step differentiation.

The numerical evidence supports four claims. The spectral methods for equal and general SLW and the Laplace-transform method agree near rounding error in the reported tests. The method retains its accuracy at volatility interfaces, where the tested uniform FD grids drop to first order and central differences do not recover the one-sided gamma limits. One-sided differences on the grid fitted to the volatility interfaces recover those limits. The Laplace-transform method has approximately constant cost over the tested maturities, while the eigenfunction expansion requires more terms at shorter maturities. Finally, the accrual-law calculation is about sixty times faster than the tested FD scheme with the accrual as a second state variable, with a smaller error.

In practice this means prices in milliseconds, Greeks including all layer vegas, and calibration in under a second. Volatility piecewise constant in time enters through re-expansion between the eigenbases of consecutive buckets, and retaining about twenty eigenfunctions per bucket reaches machine precision in the reported tests. The method also evaluates expected corridor variance based on either price or log-return quadratic variation.

For volatility piecewise constant in time, the multi-bucket construction covers terminal-payoff claims, including double-knock-out Calls and Puts, double no-touch contracts and terminal corridor digitals. It also covers linear range accruals, corridor swaps and corridor variance swaps based on either price or log-return quadratic variation, since their valuation requires expected accrued quantities. Nonlinear payoffs on accumulated occupation, including Calls on the accrued fraction, digital accruals and capped or floored accruals, are treated only for a single time bucket. Extending these products to multiple buckets requires the corresponding occupation-time transforms across successive buckets and remains for future work.

A further extension concerns moving interfaces, which arise in forward coordinates when the breakpoints are fixed in the spot price and the interest rate differs from the dividend yield. The oscillating transforms of \cite{ItkinLiptonMuravey2022_PhysicaA} provide a starting point for treating this problem. Early exercise also remains to be addressed in future work.

\section*{Acknowledgments}

The original idea of using GIT for pricing corridor variance swaps was expressed in our discussion with Peter Carr in the fall of 2021. The authors are delighted to close this gestalt and eventually elaborate on this idea, dedicating this paper to his memory.

\medskip
\noindent We are also grateful to Alexander Gairat for helpful discussions.

\medskip
\noindent During the preparation of this work, the authors used GPT-6 Astra and Claude Opus 5.5 for the purpose of validating mathematical proofs, reviewing Python source code, and refining the English language and phrasing of the manuscript. Following the use of these tools, the authors thoroughly reviewed, verified, and edited the output. The authors assume full responsibility for the mathematical accuracy, logical integrity of the proofs, and final content of this manuscript.

%%%%%%%%%%%%%%%%%%%%%%%%%%%%%%%%%%%%%%%%%%%%%%%%%%%%%%%%%%%%%%%%%%%%%%%%%%%%%

\printbibliography[title={References}]

\section*{Disclosure statement}

No potential conflict of interest was reported by the authors.

\section*{Funding}

No funding was received.

\section*{Disclaimer}

Opinions expressed here are the authors' own and do not represent the views of their employers. A standard disclaimer applies.

%%%%%%%%%%%%%%%%%%%%%%%%%%%%%%%%%%%%%%%%%%%%%%%%%%%%%%%%%%%%%%%%%%%%%%%%%%%%%
\appendix

\appendixpage
\numberwithin{equation}{section}
\setcounter{equation}{0}

\section{The lognormal and drifted variants}\label{app:drift}

For $d S = \vs(S)\,S\,d W$ with $\vs$ piecewise constant on breakpoints $0 < Y_0 < \dots < Y_N$, work with $Z := \log S$, so that $d Z = \vs(Z)\,d W - \frac 12 \vs^2(Z)\,d t$ with the breakpoints $\log Y_i$. The backward operator is $\frac 12 \vs^2(\partial_{zz} - \partial_z)$. The natural scale of $Z$ is $e^z$, so the domain condition at a breakpoint is continuity of $u$ and $u_z$. The ratio of drift to variance is $-\frac 12$ in every layer, so the substitution $u = e^{z/2} w$ removes the drift with one exponential factor common to all layers:
\begin{equation}\label{eq:lognormalw}
w_\theta = \tfrac 12 \vs_i^2\,w_{zz} - \tfrac 18 \vs_i^2\,w \quad \text{on layer } i, \qquad w \text{ and } w_z \text{ continuous}.
\end{equation}

After the spatial scale transformation $x = \sqrt 2 \int d z/\vs^2(z)$ this is the problem
of \cref{prop:charged} with the killing rates $q_i = \vs_i^2/8$. These rates are
not uniform, so by \cref{prop:charged} they cannot be removed from the kernels as a
global factor. They enter the tridiagonal system through $\nu_i = \sqrt{\lambda +
\vs_i^2/8}$ at no extra cost.

With $y_i$, $x_0$ and the density with killing $U^{(\bm q)}$ now built from the
layers in $z$, the price of a claim with payoff $P(S_T)$ is
\begin{equation}\label{eq:lognormalprice}
V = e^{-rT} e^{z_*/2} \int_{y_0}^{y_N} e^{-z(x)/2}\,P \big( e^{z(x)} \big)\,U^{(\bm q)}(T, x; x_0)\,d x, \qquad z_* := \log S_*,
\end{equation}
where $z(x)$ is affine on each layer, so the payoff integrals stay elementary for Calls, Puts and constant payoffs.

The standard log-return corridor variance has rate $\vs_i^2\ind_{\{i\in\mathcal I_C\}}$ on layer $i$. Its expectation follows by integrating the weighted physical transition kernel implied by \eqref{eq:lognormalprice}. More generally, for the transform $\EE[e^{-\beta Q_C(T)}P(S_T);T<\tau]$ on a single bucket, the same gauge transformation gives the total layer killing rates
\begin{equation}\label{eq:logvariancekilling}
q_i(\beta)=\frac{\vs_i^2}{8}+\beta\vs_i^2\ind_{\{i\in\mathcal I_C\}}.
\end{equation}
Use these rates in \eqref{eq:Kflux} and retain the gauge factors in \eqref{eq:lognormalprice}, omitting $e^{-rT}$ for the undiscounted transform. For $P\equiv1$, minus its derivative at $\beta=0$ equals $\EE[Q_C(T);T<\tau]$. Time inversion followed by inversion in $\beta$ gives the survival-weighted law of $Q_C(T)$ for time-homogeneous lognormal volatility. The tiled extension established in the main text concerns expected rewards. It does not establish this nonlinear-law calculation across time buckets.

This holds in the forward measure. If the dividend yield $d$ (the foreign interest rate for an exchange rate) differs from the interest rate $r$, the ratio of drift to variance becomes layer dependent and the exponential factor is no longer the same in all layers. Breakpoints fixed in the spot then move in the forward $F_t = S_te^{(r - d)(T - t)}$, which is the moving-interface problem of \cite{ItkinLiptonMuravey2022_PhysicaA}. When the volatility is specified in the forward, as is common for FX and rates, the model is driftless and the analysis applies unchanged. The same remark applies to a drift in the normal model.

We checked the reduction in two ways. With one layer, $\vs = 0.2$ and cut-offs at $\log S = \log 100 \pm 3$, the system with these killing rates prices the at-the-money Call at $T = 1$ within $5 \times 10^{-12}$ of the Black--Scholes value $7.965567455$. With eight layers, breakpoints $(60, 80, 90, 95, 100, 105, 110, 120, 150)$ and lognormal volatilities $(0.35, 0.28, 0.23, 0.20, 0.18, 0.17, 0.18, 0.22)$, it agrees with a finite volume scheme on the natural scale of $S$, with cell weights $\int d S/(\vs^2S^2)$ and extrapolated from 800 and 1600 intervals, to $3.9 \times 10^{-9}$ for the double no-touch and $1.9 \times 10^{-7}$ for the Call with strike $100$.

\section{The GIT representation of a layer}\label{app:potentials}

This appendix derives \eqref{eq:gradients} from the representation of \cite{ItkinMuraveyDBFMF}.

We use the representation obtained by the GIT method in \cite{ItkinMuraveyDBFMF} for the heat equation on a time-dependent interval. Consider
\begin{align} \label{eq:heat-moving}
\partial_{\tilde t} U (\tilde t, x) &= \partial_{xx} U(\tilde t, x), \qquad x \in [y(\tilde t), z(\tilde t)], \quad \tilde t > 0, \\
U(0, x) &= f(x), \qquad U(\tilde t, y(\tilde t)) = f^-(\tilde t), \qquad U(\tilde t, z(\tilde t)) = f^+(\tilde t). \nonumber
\end{align}
Let $l(\tilde t) := z(\tilde t) - y(\tilde t)$ denote the length of the interval. The unknown boundary fluxes are the gradients at the two endpoints. With the outward-normal convention,
\begin{equation} \label{PsiPhi_def}
J^-(\tilde t) = - \left. \frac{\partial U(\tilde t, x)}{\partial x} \right|_{x = y(\tilde t)}, \qquad J^+(\tilde t) = \left. \frac{\partial U(\tilde t, x)}{\partial x} \right|_{x = z(\tilde t)}.
\end{equation}
In terms of these fluxes, the solution admits the following semi-closed representation:
\begin{align} \label{U_final_alt}
U(\tilde t, x) &= \sum_{n = -\infty}^{\infty} \Bigg\{ \int_{y(0)}^{z(0)} U(0, \xi) \Upsilon_n(x, \tilde t \, | \, \xi, 0) \, d \xi + \int_0^{\tilde t} \left[ J^+(s) + f^+(s) z'(s) \right] \Upsilon_n(x, \tilde t \, | \, z(s), s) \, ds \\
& + \int_0^{\tilde t} \left[ J^-(s) - f^-(s) y'(s) \right] \Upsilon_n(x, \tilde t \, | \, y(s), s) \, ds \nonumber  \\
& + \int_0^{\tilde t} \left[ f^-(s) \mathcal K_n(x, \tilde t \, | \, y(s), s) - f^+(s) \mathcal K_n(x, \tilde t \, | \, z(s), s) \right] ds \Bigg\}, \nonumber \\
\Upsilon_n(x, \tilde t \, | \, \xi, s) &= \frac{1}{2 \sqrt{\pi (\tilde t - s)}} \left[ e^{- \frac{(2 n l(\tilde t) + x - \xi)^2}{4 (\tilde t - s)}} - e^{- \frac{(2 n l(\tilde t) + x + \xi - 2 y(\tilde t))^2}{4 (\tilde t - s)}} \right], \nonumber\\
\mathcal K_n(x, \tilde t \, | \, \xi, s) &= \frac{x - \xi + 2 n l(\tilde t)}{4 \sqrt{\pi (\tilde t - s)^3}} e^{- \frac{(2 n l(\tilde t) + x - \xi)^2}{4 (\tilde t - s)}} + \frac{x + \xi - 2 y(\tilde t) + 2 n l(\tilde t)}{4 \sqrt{\pi (\tilde t - s)^3}} e^{- \frac{(2 n l(\tilde t) + x + \xi - 2 y(\tilde t))^2}{4 (\tilde t - s)}}. \nonumber
\end{align}

For the layered construction of \cref{sec:potentials}, the general representation is specialized by introducing the rescaled time variable and identifying the moving boundaries with the endpoints of layer $i$:
\begin{align} \label{eq:rescaling}
\tilde t &= \sigma_i^2 t, \qquad U(\tilde t, x) = U_i(t, x), \qquad f^+(\tilde t) = f_i(t), \qquad f^-(\tilde t) = f_{i - 1}(t), \\
y(\tilde t) &= y_{i - 1}, \qquad z(\tilde t) = y_i, \qquad l(\tilde t) = \ell_i = y_i - y_{i - 1}. \nonumber
\end{align}

After this rescaling, the layer solution takes the form
\begin{align} \label{eq:Ufinal_alt1}
U_i(t, x) &= \sum_{n = -\infty}^{\infty} \sigma_i^2 \Bigg\{ \int_0^t J^+(\sigma_i^2 s) \Upsilon_n(x, \sigma_i^2 t \, | \, y_i, \sigma_i^2 s) \, ds + \int_0^t J^-(\sigma_i^2 s) \Upsilon_n(x, \sigma_i^2 t \, | \, y_{i - 1}, \sigma_i^2 s) \, ds \nonumber\\
&+ \int_0^t \left[ f_{i - 1}(s) \mathcal K_n(x, \sigma_i^2 t \, | \, y_{i - 1}, \sigma_i^2 s) - f_i(s) \mathcal K_n(x, \sigma_i^2 t \, | \, y_i, \sigma_i^2 s) \right] ds \Bigg\}.
\end{align}

To avoid carrying the rescaled time arguments throughout the subsequent calculations, we redefine the kernels in terms of the original time variable:
\begin{align} \label{eq:rescaled-kernels}
\Upsilon_n(x, t \, | \, \xi, s) &= \Upsilon_n(x, \sigma_i^2 t \, | \, \xi, \sigma_i^2 s) = \frac{1}{2 \sigma_i \sqrt{\pi (t - s)}} \left[ e^{- \frac{(2 n \ell_i + x - \xi)^2}{4 \sigma_i^2 (t - s)}} - e^{- \frac{(2 n \ell_i + x + \xi - 2 y_{i - 1})^2}{4 \sigma_i^2 (t - s)}} \right], \\
\mathcal K_n(x, t \, | \, \xi, s) &= \mathcal K_n(x, \sigma_i^2 t \, | \, \xi, \sigma_i^2 s) = \frac{x - \xi + 2 n \ell_i}{4 \sigma_i^3 \sqrt{\pi (t - s)^3}} e^{- \frac{(2 n \ell_i + x - \xi)^2}{4 \sigma_i^2 (t - s)}} + \frac{x + \xi - 2 y_{i - 1} + 2 n \ell_i}{4 \sigma_i^3 \sqrt{\pi (t - s)^3}} e^{- \frac{(2 n \ell_i + x + \xi - 2 y_{i - 1})^2}{4 \sigma_i^2 (t - s)}}, \nonumber
\end{align}
as well as the gradients:
\begin{eqnarray}
J^-_i(t) := J^-(\sigma_i^2 t) = J^-(\tilde t) = - \left. \frac{\partial U(\tilde t, x)}{\partial x} \right|_{x = y(\tilde t)} = - \left. \frac{\partial U_i(t, x)}{\partial x} \right|_{x = y_{i - 1}}, \nonumber\\
J^+_i(t) := J^+(\sigma_i^2 t) = J^+(\tilde t) = \left. \frac{\partial U(\tilde t, x)}{\partial x} \right|_{x = z(\tilde t)} = \left. \frac{\partial U_i(t, x)}{\partial x} \right|_{x = y_i}
\end{eqnarray}
Combining all terms together yields the formula for $U_i(t, x)$
\begin{align} \label{eq:Ufinal_alt2}
U_i(t, x) &= \sum_{n = -\infty}^{\infty} \sigma_i^2 \Bigg\{ \int_0^t J^+_i(s) \Upsilon_n(x, t \, | \, y_i,  s) \, ds + \int_0^t J^-_i(s) \Upsilon_n(x, t \, | \, y_{i - 1}, s) \, ds \nonumber\\
&+ \int_0^t \left[ f_{i - 1}(s) \mathcal K_n(x, t \, | \, y_{i - 1}, s) - f_i(s) \mathcal K_n(x,  t \, | \, y_i, s) \right] ds \Bigg\}.
\end{align}
Differentiating \eqref{eq:Ufinal_alt2} with respect to $x$ and evaluating at the boundary yields the following system of Volterra equations of the second kind:
\begin{align} \label{eq:VolterraEq}
-J^-_i(t) &= \sum_{n = -\infty}^{\infty} \sigma_i^2 \Bigg\{ \int_0^t J^+_i(s)  \frac{\partial \Upsilon_n(x, t \, | \, y_i,  s)}{\partial x} \bigg|_{x = y_{i - 1}} \, ds + \int_0^t J^-_i(s) \frac{\partial \Upsilon_n(x, t \, | \, y_{i - 1}, s)}{\partial x } \bigg|_{x = y_{i - 1}}  \, ds \nonumber\\
&+ \int_0^t \left[ f_{i - 1}(s) \frac{\partial \mathcal K_n(x, t \, | \, y_{i - 1}, s)}{\partial x} \bigg|_{x = y_{i - 1}} - f_i(s)  \frac{\partial \mathcal K_n(x,  t \, | \, y_i, s)}{\partial x } \bigg|_{x = y_{i - 1}} \right] ds \Bigg\}, \\
J^+_i(t) &= \sum_{n = -\infty}^{\infty} \sigma_i^2 \Bigg\{ \int_0^t J^+_i(s)  \frac{\partial \Upsilon_n(x, t \, | \, y_i,  s)}{\partial x} \bigg|_{x = y_{i}} \, ds + \int_0^t J^-_i(s) \frac{\partial \Upsilon_n(x, t \, | \, y_{i - 1}, s)}{\partial x } \bigg|_{x = y_{i}}  \, ds \nonumber\\
&+ \int_0^t \left[ f_{i - 1}(s) \frac{\partial \mathcal K_n(x, t \, | \, y_{i - 1}, s)}{\partial x} \bigg|_{x = y_{i}} - f_i(s)  \frac{\partial \mathcal K_n(x,  t \, | \, y_i, s)}{\partial x } \bigg|_{x = y_{i}} \right] ds \Bigg\}.
\end{align}
Some parts of the last expression simplify, in particular the terms containing $\Upsilon_n$. Interchanging summation and integration and evaluating at the boundary values yields

\noindent
\resizebox{\textwidth}{!}{%
\begin{minipage}{\linewidth}
\begin{align*}
&\sum_{n = -\infty}^{\infty} \frac{\partial \Upsilon_n(x, t \, | \, \xi,  s)}{\partial x} = -\frac{1}{4 \sigma_i^3 \sqrt{\pi (t - s)^3}} \sum_{n = -\infty}^{\infty} \left[ (2 n \ell_i + x - \xi) e^{- \frac{(2 n \ell_i + x - \xi)^2}{4 \sigma_i^2 (t - s)}} - (2 n \ell_i + x + \xi - 2 y_{i - 1}) e^{- \frac{(2 n \ell_i + x + \xi - 2 y_{i - 1})^2}{4 \sigma_i^2 (t - s)}} \right],
\end{align*}
\end{minipage}
}

\medskip
\noindent and

\noindent
\resizebox{\textwidth}{!}{%
\begin{minipage}{\linewidth}
\begin{align*}
\sum_{n = -\infty}^{\infty} \frac{\partial \Upsilon_n(x, t \, | \, y_{i - 1},  s)}{\partial x} \bigg|_{x = y_{i - 1}} &=
-\frac{1}{4 \sigma_i^3 \sqrt{\pi (t - s)^3}} \sum_{n = -\infty}^{\infty} \left[ (2 n \ell_i) e^{- \frac{(2 n \ell_i )^2}{4 \sigma_i^2 (t - s)}} - (2 n \ell_i) e^{- \frac{(2 n \ell_i)^2}{4 \sigma_i^2 (t - s)}} \right] = 0, \\
\sum_{n = -\infty}^{\infty} \frac{\partial \Upsilon_n(x, t \, | \, y_{i - 1},  s)}{\partial x} \bigg|_{x = y_{i}} &= -\frac{1}{4 \sigma_i^3 \sqrt{\pi (t - s)^3}} \sum_{n = -\infty}^{\infty} \left[ ((2 n + 1) \ell_i) e^{- \frac{((2 n + 1) \ell_i)^2}{4 \sigma_i^2 (t - s)}} - ((2 n + 1) \ell_i) e^{- \frac{((2 n + 1) \ell_i )^2}{4 \sigma_i^2 (t - s)}} \right] = 0, \\
\sum_{n = -\infty}^{\infty} \frac{\partial \Upsilon_n(x, t \, | \, y_i,  s)}{\partial x} \bigg|_{x = y_{i - 1}}
&=
-\frac{1}{4 \sigma_i^3 \sqrt{\pi (t - s)^3}} \sum_{n = -\infty}^{\infty} \left[ ((2 n - 1) \ell_i) e^{- \frac{((2 n - 1) \ell_i )^2}{4 \sigma_i^2 (t - s)}} - ((2n - 1) \ell_i ) e^{- \frac{((2 n - 1) \ell_i )^2}{4 \sigma_i^2 (t - s)}} \right] = 0, \\
\sum_{n = -\infty}^{\infty} \frac{\partial \Upsilon_n(x, t \, | \, y_i,  s)}{\partial x} \bigg|_{x = y_{i}}
&=
-\frac{1}{4 \sigma_i^3 \sqrt{\pi (t - s)^3}} \sum_{n = -\infty}^{\infty} \left[ (2 n \ell_i) e^{- \frac{(2 n \ell_i)^2}{4 \sigma_i^2 (t - s)}} - ( (2 n + 2) \ell_i) e^{- \frac{((2 n + 2) \ell_i )^2}{4 \sigma_i^2 (t - s)}} \right] = 0.
\end{align*}
\end{minipage}
}

\medskip
\noindent Therefore, the Volterra equations reduce to explicit expressions for the gradients $J^-_i(t)$ and $J^+_i(t)$:
\begin{align} \label{eq:VolterraEq2}
-J^-_i(t) &= \sum_{n = -\infty}^{\infty} \sigma_i^2 \int_0^t \left[ f_{i - 1}(s) \frac{\partial \mathcal K_n(x, t \, | \, y_{i - 1}, s)}{\partial x} \bigg|_{x = y_{i - 1}} - f_i(s)  \frac{\partial \mathcal K_n(x,  t \, | \, y_i, s)}{\partial x } \bigg|_{x = y_{i - 1}} \right] ds, \\
J^+_i(t) &= \sum_{n = -\infty}^{\infty} \sigma_i^2 \int_0^t \left[ f_{i - 1}(s) \frac{\partial \mathcal K_n(x, t \, | \, y_{i - 1}, s)}{\partial x} \bigg|_{x = y_{i}} - f_i(s)  \frac{\partial \mathcal K_n(x,  t \, | \, y_i, s)}{\partial x } \bigg|_{x = y_{i}} \right] ds. \nonumber
\end{align}

\noindent For the terms containing $\mathcal K_n(x,  t \, | \, \xi, s)$ we may also do some simplifications. Indeed,
\begin{align*}
\sigma_i^2 \frac{\partial \mathcal K_n(x, t \, | \, \xi, s)}{\partial x } &= \frac{1}{4 \sigma_i \sqrt{\pi (t - s)^3}} e^{- \frac{(2 n \ell_i + x - \xi)^2}{4 \sigma_i^2 (t - s)}} + \frac{1}{4 \sigma_i \sqrt{\pi (t - s)^3}} e^{- \frac{(2 n \ell_i + x + \xi - 2 y_{i - 1})^2}{4 \sigma_i^2 (t - s)}} \\
 &- \frac{(x - \xi + 2 n \ell_i)^2}{8 \sigma_i^3 \sqrt{\pi (t - s)^5}} e^{- \frac{(2 n \ell_i + x - \xi)^2}{4 \sigma_i^2 (t - s)}} - \frac{(x + \xi - 2 y_{i - 1} + 2 n \ell_i)^2}{8 \sigma_i^3 \sqrt{\pi (t - s)^5}} e^{- \frac{(2 n \ell_i + x + \xi - 2 y_{i - 1})^2}{4 \sigma_i^2 (t - s)}} \nonumber \\
 &= \frac{\partial }{\partial s} \left( \frac{1}{2 \sigma_i \sqrt{\pi (t - s)}} e^{- \frac{(2 n \ell_i + x - \xi)^2}{4 \sigma_i^2 (t - s)}} + \frac{1}{2 \sigma_i \sqrt{\pi (t - s)}} e^{- \frac{(2 n \ell_i + x + \xi - 2 y_{i - 1})^2}{4 \sigma_i^2 (t - s)}} \right).
\end{align*}
Hence,
\begin{align*}
\sigma_i^2 \sum_{n = -\infty}^{\infty} \frac{\partial \mathcal K_n(x, t \, | \, y_{i - 1}, s)}{\partial x} \bigg|_{x = y_{i - 1}} &= \sigma_i^2 \sum_{n = -\infty}^{\infty} \frac{\partial \mathcal K_n(x, t \, | \, y_{i}, s)}{\partial x} \bigg|_{x = y_{i}} =   \frac{\partial } {\partial s} \left( \sum_{n = -\infty}^{\infty} \frac{1}{\sigma_i \sqrt{\pi (t - s)}} e^{- \frac{(2 n \ell_i)^2}{4 \sigma_i^2 (t - s)}} \right), \\
\sigma_i^2 \sum_{n = -\infty}^{\infty} \frac{\partial \mathcal K_n(x, t \, | \, y_{i - 1}, s)}{\partial x} \bigg|_{x = y_{i}} &= \sigma_i^2 \sum_{n = -\infty}^{\infty} \frac{\partial \mathcal K_n(x, t \, | \, y_{i}, s)}{\partial x} \bigg|_{x = y_{i - 1}} = \frac{\partial }{\partial s} \left( \sum_{n = -\infty}^{\infty}  \frac{1}{\sigma_i \sqrt{\pi (t - s)}} e^{- \frac{((2 n + 1) \ell_i)^2}{4 \sigma_i^2 (t - s)}} \right).
\end{align*}
With the image kernels \eqref{eq:kernels}, an integration by parts turns these relations into \eqref{eq:gradients}.

The convolution form \eqref{eq:gradients} has a one-line proof in the Laplace domain. On layer $i$ the transformed layer solution with end values $\Lap f_{i - 1}$ and $\Lap f_i$ is
\[
\Lap U_i(\lambda, x) = \frac{\Lap f_{i - 1}\,\sinh \big( \sqrt \lambda(y_i - x)/\sigma_i \big) + \Lap f_i\,\sinh \big( \sqrt \lambda(x - y_{i - 1})/\sigma_i \big)}{\sinh(\omega_i \sqrt \lambda)}.
\]
Its gradient at $y_{i - 1}$ is
\[
(\sqrt \lambda/\sigma_i) \big( \Lap f_i/\sinh(\omega_i \sqrt \lambda) - \Lap f_{i - 1} \coth(\omega_i \sqrt \lambda) \big).
\]
By \eqref{eq:Letas} and $\Lap f' = \lambda \Lap f$, this is the transform of $-\etaev{i}*f'_{i - 1} + \etaod{i}*f'_i$, which is the first line of \eqref{eq:gradients}. The same computation at $y_i$ gives the second line.

In the time domain the derivation above passes through hypersingular integrals. Near $s = t$ the kernel $\partial_s \etaev{i}(t - s)$ behaves like $(t - s)^{-3/2}$, so the integration by parts must be read in the finite-part sense. The divergent boundary terms then cancel, and the lower boundary term vanishes because $f_i(0) = 0$. The Laplace computation avoids these steps, and it is the proof we rely on.

\section{Layer-dependent killing: algebra}\label{app:charged}

On layer $m$ the transform of the density with killing is a combination of $e^{\pm \nu_m(x - y_{m - 1})/\sigma_m}$. In computations every term carries a nonpositive real exponent. With $\kappa := \nu_m/\sigma_m$, $\Re \kappa \ge 0$, $s := x - y_{m - 1} \in[0, \ell_m]$ and $D := 1 - e^{-2 \kappa \ell_m}$,
\[
\frac{\sinh(\kappa(\ell_m - s))}{\sinh(\kappa \ell_m)} = \frac{e^{-\kappa s} - e^{-\kappa(2 \ell_m - s)}}{D},   \qquad \frac{\sinh(\kappa s)}{\sinh(\kappa \ell_m)} = \frac{e^{-\kappa(\ell_m - s)} - e^{-\kappa(\ell_m + s)}}{D}.
\]

On the source layer, with $s_0 := x_0 - y_{j - 1}$ and $\kappa := \nu_j/\sigma_j$,
\[
\Lap H_j^{(\bm q)} = \frac{e^{-\kappa|s - s_0|} - e^{-\kappa(s + s_0)} - e^{-\kappa(2 \ell_j - s - s_0)} + e^{-\kappa(2 \ell_j - |s - s_0|)}}{2\,\sigma_j \nu_j\,(1 - e^{-2 \kappa \ell_j})},
\]
which is the method of images in the Laplace domain. Payoff integrals reduce to
$\int(\alpha + \beta s)\,e^{p + cs}\,d s$ in closed form, with a Taylor series when
$|c|$ times the interval length is below $\frac 12$.

For the inversion in the accrual we use the rectangle of \cref{sec:invert} with $\bar c = 1/T$ and $\delta = 2/T$, horizontal edges truncated at $\Re \lambda = \bar c - 40/T$, and Gauss--Legendre panels with 16 nodes, of width $\min(\pi/T, 1)$ on the right edge and $\min(1/T, 1)$ on the horizontal edges. All nodes of one contour go through a single vectorized solution of the tridiagonal systems. Real killing rates need no rectangle, because their singularities lie on the negative real axis and the Talbot rule applies.

\section{Eigen-solver details}\label{app:eigen}

\myparagraph{Counting and bisection.} The count \eqref{eq:count} needs
$\Kmat(\xi)$ only through its diagonal and off-diagonal, evaluated with $\cot$ and $1/\sin$. The count is evaluated away from the poles $\omega_i\xi\in\pi\mathbb Z$. Near a pole, large but finite floating-point entries do not guarantee a correct inertia count. The trial point must be moved within its bracket or the calculation repeated at higher precision. A root coinciding with a layer pole is bracketed by counts on either side. For all modes below $\xi_{\max}$ we run the bisection simultaneously. Mode $n$ keeps a bracket and moves the end on the side where the count at the midpoint is below $n$ or at least $n$. Deflation is never needed, because modes are located by index.

For the inertia computation, a nonsingular matrix may have a zero leading scalar pivot. Symmetric $LDL^{\top}$ factorization with $1\times1$ and $2\times2$ pivot blocks handles this case. The number of negative eigenvalues of the blocks gives the inertia. Unresolved pivots require moving the trial point or higher precision.

\myparagraph{Eigenfunctions.} Without killing we propagate the pair
$(u, \sigma^2u_x)$ from $(0, 1)$ at $y_0$ by \eqref{eq:transfer}. As explained in \cref{rem:transferconditioning}, layer rotations are separated by interface rescalings. For large contrasts or many layers, propagation can be stabilized by renormalizing the pair while retaining its accumulated scale, or by using a phase and logarithmic amplitude. The eigenvalues of the general method agree with the closed form of
\cref{thm:equalmodes} to a relative $3.8 \times 10^{-15}$ on the equal-SLW
configuration.

\myparagraph{Vegas.} For \eqref{eq:vega} we shoot in the price coordinate, where the geometry does not depend on $\vs$. On layer $m$, with $s := y - Y_{m - 1}$ and $k_m := \sqrt{2 \mu}/\vs_m$, the solution is a linear combination of $\cos(k_m s)$ and $\sin(k_m s)$, the pair $(u, u_y)$ is continuous across interfaces, and the shooting function is $\mathcal D(\mu, \vs) := u(B_+)$ for the initial pair $(0, 1)$. We propagate $\partial_\mu u$ and $\partial_\mu u_y$ alongside, which gives $\partial_\mu \mathcal D$ exactly. Let $\bm e_i$ be the $i$-th coordinate vector. For each layer $i$ we continue every root to the complex volatility vector $\vs + \mathrm{i} h \bm e_i$ with $h = 10^{-20}$ by two Newton steps on $\mathcal D$ from the real root, evaluate the series in complex arithmetic, and take $\Im V/h$. The imaginary parts of the continued roots reproduce \eqref{eq:HF} to rounding.

Complex-step differentiation has truncation error $O(h^2)$ for analytic expressions and requires the imaginary signal of size $h$ to be preserved throughout the computation. Conjugation, absolute values and nonanalytic root selection must be avoided on perturbed data. Norm formulas are continued through analytic bilinear expressions, and mode labels and spectral cutoffs are held fixed. Writing the payoff integrals
$\int(\alpha + \beta s) \cos(ks)\,d s$ through $e^{\pm \mathrm{i} ks}$ leaves imaginary
rounding noise of order $10^{-16}$ in a real quantity, which a step of $10^{-20}$
cannot survive. We therefore evaluate them with real trigonometric functions.

\end{document}